\documentstyle[11pt,psfig]{article}
\begin{document}
 
\vspace{-4cm}

\title
{
Evolution and extinction dynamics
 in rugged fitness landscapes.
}
\author{Paolo Sibani and Michael Brandt\\
Fysisk Institut\\
 Odense Universitet,\\
  Campusvej 55, DK5230 Odense M\\
  Denmark
 \and
Preben Alstr\o m\\
Niels Bohrs Institut\\
K\o benhavns Universitet,\\
Blegdamvej 17, DK2100 K\o benhavn N\\
Denmark
}
\maketitle
\vspace{-1cm}
\begin{abstract}
\begin{small}

After an introductory section summarizing 
the paleontological data and  some of 
 their  theoretical descriptions, 
we  describe  the `reset'
model and its (in part analytically soluble) mean field  
 version, which have been  briefly introduced    in
  Letters\cite{Sibani95,Sibani97}.
Macroevolution is considered as a problem of stochastic dynamics in a system 
with many competing agents. Evolutionary events (speciations and 
extinctions) are triggered by fitness records found by random
exploration of the agents' fitness landscapes. As a  consequence,
the average fitness in the system increases logarithmically with time, while  
the rate of extinction steadily decreases. This  non-stationary  dynamics is  studied  
by numerical simulations and, in a simpler mean field version, analytically.
We also consider  the effect of externally added `mass' extinctions.  
The predictions for various quantities of paleontological interest (life-time
distribution, distribution of event sizes and behavior of the rate of
extinction) are robust and in good agreement with  available data.  
\end{small}
\end{abstract}
\noindent

\section{Introduction and background} 
Life on earth originated 3.5 billion years ago. According to the earliest
record of life, the first organisms were primitive one-celled
non-photosynthesizing bacteria. Two billion years then followed, before
more complex multicelled organisms appeared in a catastrophic
event called the Cambrian explosion. This evolutionary event, which took
place about 600 million years ago, led to a diversity of new species,
and to  a remarkable fossil record for the following Phanerozoic
period, which provides  valuable and fundamental material for scientists who
seek insight into the fascinating history of life on earth.

Unfortunately, the fossil record is often limited when it comes to
understanding the development of a given species at the 
level of organisms, i.e.  the
micro-evolutionary processes. However, the record makes out a good
statistical material at the macro-evolutionary level, where species
and higher taxa are the  basic units considered\cite{Sepkoski93}.
At this  and higher taxonomical levels,  evolutionary
measures,  as for instance the distribution of life spans, 
provide a good characterization of  biological evolution
and a basis for understanding its underlying mechanism.  
 
The evolution of species is typically sketched in the form of evolutionary
trees   with several lineage branchings. Whenever a lineage
branches, it marks the appearance of new species (speciation), and
whenever it  stops, it marks the extinction of a species.
We note that almost all species are extinct\cite{Raup95}. The
possibility of hybridization, where two lineages merge to form a new
intermediate species, is normally not seen.
Other evolutionary changes are essential, the most prominent being
the phyletic transformations taking place along a lineage.
These transformations are basically Darwinian changes in species
due to   natural selection. The
extinctions associated with phyletic transformations are called
pseudo-extinctions. 

Several questions arise  when this Darwinian picture of evolution 
is taken under closer scrutiny.  Did species always have time to adapt 
to the changes in their environment?  And  what are the
selection mechanisms ( if any ) that on a species level 
determine  who is to survive and who is to go extinct?  
Is evolution mostly characterized by gradual changes, or is
it rather characterized by `quantal jumps' or `punctuations',
where new species are formed on a relatively short geological 
time scale?  

We may further ask: can evolution be described by    
  dynamical processes fluctuating around fix  
values,  or is there a sense of direction in biological evolution,
 expressed, for example,  by  a steadily  increasing fitness measure?
Why should nature select
an average life time for species to be four million years? Why not four thousand
years or 4 billion years?

Explanations of evolutionary and extinction history 
can be coarsely   classified  
according to their standing on the question of 
 gradualism, to the importance they confer to    
  `bad luck' as opposed to  `bad genes'\cite{Raup91}, 
 and finally according to their view on 
 stationarity. All these issue 
  have   been  intensely debated by paleontologist.
  From fossil data, it seems to us that nature has indeed
not selected an equilibrium. For example, the standing diversity 
of families of marine vertebrates\cite{Raup82} has increased by
a large factor  over the
last 600 million years while, in the same period,  
  the body plan complexity    ( the number
of different somatic cells in the most complex of the taxa
extant at a given time )  has increased  sixhundredfold\cite{Valentine95}. 
The extinction rate does not appear to have  
reached an equilibrium either. Rather  
it seems to have a downward trend,   as   noted by Raup and
Sepkoski\cite{Raup82} in 1982. Their data are reproduced in 
Fig.~1a. The decay of extinction intensity
 can be highlighted  by integrating the extinction rate, measured as a percent of 
standing diversity. The accumulated extinction rate percent describes the
number of extinctions in a (hypothetical) system with a   constant
standing diversity. If evolutionary dynamics were a stationary process,
with -- on average -- an equal number of events taking place in any 
given time interval, this quantity would increase linearly in time.
Instead, as shown in Fig.~1b, it increases as a power law $R(t) = t^{1-\gamma}$,
and, correspondingly, the percentual rate of extinction $r(t)$ decays  as $t^{-\gamma}$.
The numerical value of  $\gamma$ is  $\approx 0.4$. 
Admittedly, the extinction record   shows  large fluctuations.
The  events  which fall  outside the expected
statistical variation\cite{Raup_disagrees}  are mainly related to  the five
big mass extinctions, which  have attracted enormous
interest, presumably due to their catastrophic impact on
the ecosystem. 
The now most commonly accepted  explanation of these events  are
 meteor impacts\cite{Alvarez80}.
\clearpage  
\begin{figure}[pt]
 \centerline{\psfig{figure=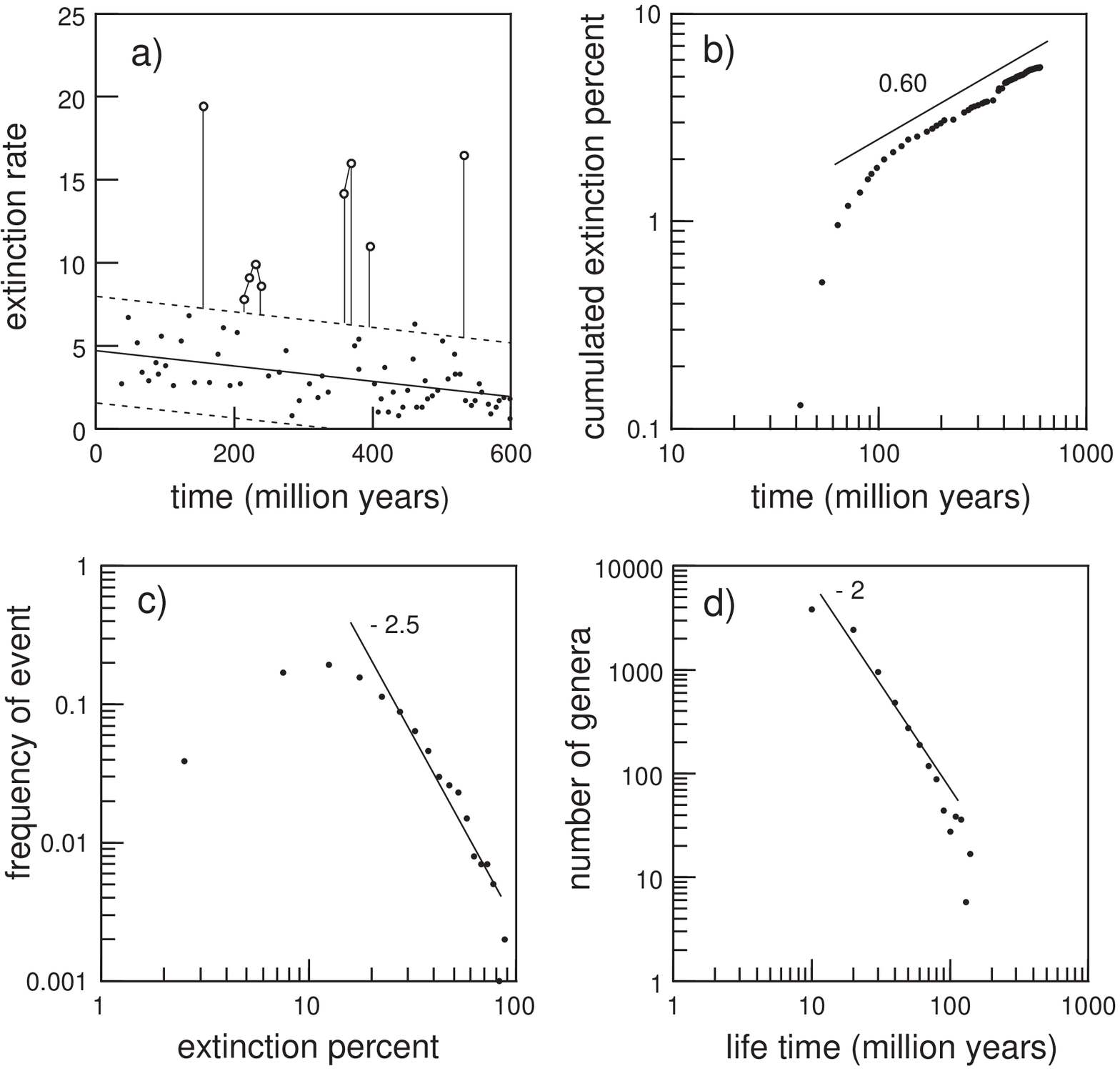,height=12.5cm,width=14cm}}  
  \vfill 
\begin{caption} {}
\begin{scriptsize} 
 a) The extinction rate (extinction per million years)
 for families of marine vertebrates and invertebrates  is plotted as
 a function of geological time. These data are redrawn from
 Ref.\cite{Raup82}. b) Using the  
 information on  the corresponding  standing diversity 
 available from the same source, the extinction rate
  is converted in an extinction percentage, which is then
  integrated and   shown in a log-log plot as a 
  function of    time. If extinction were 
  a stationary process, this cumulative extinction 
  percentage  should fall on a line with slope 
  $1$. Instead,  as highlighted in the plot, the majority of data points fall 
  on  a line with slope $0.6$. In plots a) and b) the
  origin of time is set to $600$ million years ago, corresponding to the
  beginning of the Phanerozoic period. c) The shown   data describe the
  distribution of extinction event sizes, expressed as fraction of the
  extant species killed. The data are redrawn  from Ref.~\cite{Raup95}.
  As shown in the figure the distibution of event sizes falls off
  approximately as a power law. The exponent is $-2.5$, which, as we
  argue in the text, is the reciprocal of the exponent characterizing 
  the decay of the rate of extinction. 
  \end{scriptsize} 
\end{caption}
\end{figure}
\newpage  
Meteor craters have been looked for and identified to account not
only for the mass extinctions but also for several minor events, where
a diversity of   species  seems to suddenly disappear within a
short geological time span. However, since   mass extinctions only 
represent a small fraction of the data,
 they only exert a relatively minor influence on the extinction rate. 

An intriguing result of the decay of the extinction rate is its direct
influence   on other evolutionary measures.
Consider for instance the frequency $G(x)$ of extinction events
in which a fraction $x $ of the extant species are killed. The empirical form
of this quantity  is reproduced   in Fig.1c   after Raup's data \cite{Raup95}.
If $p(x,t)$ is  defined as the joint  probability density 
that an extinction of size $x$ happens at time $t$, we then have 
  $ G(x) \propto \int p(x,t) dt $
and $r(t) \propto  \int p(x,t) dx $. Let us now for simplicity
 neglect  fluctuations altogether,  i.e. assume that 
extinctions of a given size   only take place  
at a given time, with the biggest extinction coming  first.
If a stage has length $d$, the fraction $x_0(t)$ of the system  which is removed at 
time $t$ is  proportional to $ d r(t)$. Furthermore, one can easily  show that
 our deterministic assumption, together with the form of the 
marginal distributions, implies $p(x,t) \propto  \delta(x-x_0(t)) x$.
We then  find by integrating over   $t$ that $ G(x) \propto x^{-1/\gamma}$.
In other words  this ( undoubtedly oversimplified )  argument
 predicts that the size distribution should decay as a power-law,
 with an exponent which is the reciprocal of the one characterizing the decay of the 
rate of extinction. A glance to Fig.1c shows that - in spite of the 
approximation involved,   
the prediction is in reasonable agreement with the data, as far as the 
decaying part of the curve is concerned. The fact that we are `missing' 
some of the small   extinction events is -- in this interpretation --
attributed to  a `finite time' effect, in that
 the very small   events are yet  to come. An additional  source
 of discrepancy pulling in the same direction  is   that small events might be
 simply  lost  in the noise.
This interpretation differs from  the stationary view of evolution 
implied in the kill curve\cite{Raup92}, where   events of any size can happen
 with a time independent  probability distribution.  
By way of contrast,  in a non-stationary scenario,  the
mean waiting time for an event of a given size,
together  with  other statistical parameters,
  strongly depends  on the    age of the system.

A last  very  important available evolutionary measure is the distribution of
the life-time 
of species, which is shown in Fig.~1d. These data were tabulated by   Raup\cite{Raup95}
based on the compilation of Sepkoski\cite{Sepkoski93}. They
 describe the empirical life-time
distribution  of about $17 500$ extinct genera of marine animals. They cover about $100$
million years and display a very clean $t^{-2}$ dependence. The distribution lacks
an average,  a wideness which   matches
 the variation in the  life times of now existing species.
 
\section{Evolution and extinctions models} 
Available evolutionary data present a challenge  for the 
theoretician, a challenge which has been recently taken
up in  several different approaches
\cite{Sibani95,Sibani97,Bak93,Sneppen95,Newman95,Newman97,Sneppen97,Sole96,Manrubia97}.
Starting  from    extremely 
simplified  assumptions,   these models    make use of 
 ideas and techniques borrowed form physics,
particularly the statistical mechanics of interacting systems,
in order to find    quantitative predictions for evolutionary measures.
It seems unlikely that any  single   model should capture all the 
complexity  of biological evolution. An additional reason to keep an open mind on
the issue of modeling is that even the interpretation of rather basic data 
is still open  to discussion.  It has for instance been suggested\cite{Raup84,Sepkoski89}
that   extinction events should be periodic, with a period of approximately
$26$ million years.  The cause of periodicity 
is usually assumed to be  external forcing. By way of contrast,
 other   descriptions - including the present one  -  
view  extinction data as the  realization of a stochastic process,
where   peaks and valleys  do not need a detailed
explanation. Instead, one is interested in understanding  the distribution
which generates  the  events.
   
Because  evolution on a large scale is hardly a reproducible
experiment, one might pessimistically believe that   
 modeling beyond data fitting  is a pointless exercise.    
Our point of view is that -- on the contrary --
  exploring in detail the  mathematical  consequences of
various assumptions  will eventually contribute
to a clarification.
  
Models strongly differ on the issue of  the relationship between
evolution and extinction. In some cases
 \cite{Sneppen95,Sibani95,Sibani97}  it is assumed
that the evolution of one species affects the
likelihood that  a   `neighbor' species would evolve.
Evolutionary landscapes of different species 
become then linked to one  another through  ecological
 interactions. Modifications of the abiotic environment
 exemplified by meteor impacts or any
other source of stress  are    included as an
additional mechanism in Ref.~\cite{Newman95}, while 
in Ref.~\cite{Newman97} they  are   considered the
exclusive   causes of   extinctions in a system of non-interacting 
species.

The approximate  scale invariance  in  several  evolutionary
measures (see Fig.~1) has prompted the hypothesis  that 
evolutionary systems should be in a
  `self organized critical state'.  The Bak and Sneppen model
  which explores this idea is extremely  simple:
 A set of agents is placed on a line. Each agent - or species - is characterized
 by one random number,  a  barrier which has to be overcome.
At any given time the   agent with the lowest barrier evolves, i.e. it receives 
a new {\em random}
number. The evolution of one agent   triggers the movement of its neighbors,
 which are then removed from the lattice (evolve or die) and   replaced 
 by new agents with a random fitness. With this dynamical prescription,
the system organizes itself into a state of dynamical equilibrium,
where the overwhelming majority of the agents has fitness above 
a well defined  critical threshold.
An `event' or avalanche starts when a fluctuations pushes at least one 
 agent  below the thresholds and subsides when the system  returns to normality.
The size distribution of the  avalanches is a power-law   with an exponent close
to $-1.1$. This is   relatively far from the behavior of the data,
 which are better described by an exponent close to   $-2.5$ (see Fig. 1c).   
The  distribution of  the life-times  of the agents is likewise a power-law,
 with an exponent $-1.1$, also
  relatively far from the correct value of $-2$.  
Newman and Roberts\cite{Newman95} have generalized  the Bak-Sneppen model by including
the effect of environmental forces, which are modeled by a randomly distributed
fluctuating  stress.
Agents with fitness lower than the current   stress value die and are replaced by new agents,
whose fitness is randomly assigned. This model has the advantage over the Bak-Sneppen model
of more clearly  distinguishing between evolution and extinctions.
It likewise  offers a scale invariant event size distribution,  with 
an exponent close to $-2$ (and closer to the data).   In a later paper\cite{Newman97} Newman
considers  a different   model,
where the  direct co-evolutionary element is removed,
 while  the  effect of the external 
environment remains the one just described.    This is in other words a 
pure `bad luck' model, with non interacting agents subject to
a common source of  external stresses, also referred to as  `coherent noise'.
 The distribution of the event sizes is only very weakly dependent on the
type of  noise,  and is in most cases well characterized by a power-law with an exponent
$-2$ over a wide range of scales. The distribution of life-times is   a power-law with
an  exponent close to $-1$.  A  related paper by
Sneppen and Newman\cite{Sneppen97} analyzes    this model  
from a more technical point  of view.
The authors find that for a wide variety 
of noise distributions the event sizes are 
distributed as $s^{-x}$, with 
$x = 1 + \alpha$ and  $\alpha \approx 1$,
 slightly depending on the noise distribution. 
The life-time distribution is found to have the power-law form
 $t^{-2 + 1/\alpha}$. Therefore, it does not seem possible
in  this  approach to have 
  exponents  for the size distribution  and
the  life-time distribution  which  both are  in the correct 
range.  

The model of Sol\'{e} and Manrubia\cite{Sole96} and an
 analytically tractable further  development by 
Manrubia and Paczuski\cite{Manrubia97}
strongly emphasize ecological interactions among species,
and  do  not include the possibility of evolution of  species
in isolation.  The interactions  
are defined by a connection matrix $W$, which has positive
as well as   negative entries. The sum of the elements of $W$  in column
$i$ defines a `viability' $v(i) $ of species $i$.  The species goes 
extinct when its viability  goes below zero. It is then replaced by a speciation 
process, in which  the new agent filling the niche 
strongly resembles one of the   extant species. (Which  
differs from all the other cases described, where   replacements 
typically occur at
random).  The dynamics is
driven by random changes in the connection matrix and leads to a
power-law distribution of extinction events with an exponent close
to $-2$. In the simplified analytical version the decrease in    
viability is  put into   the model rather than following from   
the change of the connectivity matrix.  This model   predicts 
the correct  form of 
the  life time distribution and of the extinction event size.
Being stationary, it also  has a constant average rate of extinction.
On the other hand it   produces an endogenous oscillatory 
behavior (waves of extinctions) which could  offer 
an explanation of the possible periodicity of extinctions.  
 
Most authors -- implicitely or explicitly --
 assume that evolution is a fluctuation driven process.
In this paper we take the opposite view, 
expanding     a non-stationary model  of macroevolution
which has previously been  described in two brief 
communications\cite{Sibani95,Sibani97}. We provide more detail
on the  model itself and a more complete discussion of its
relationship to the empirical data and to other approaches.

The fact that  a slow non-trivial transient dynamics is 
present in  biological evolution seems to us a  clear 
feature of the  data  which 
calls for further  studies of the fossil record from a temporal
perspective. We also note  that a  slow transient dynamics seems  
 a general inherent property of strongly interacting systems
  with many degrees of freedom and a complex state space. 
 
In the next sections, we shall pursue the idea of a transient evolution
dynamics. Mathematical results that these  
models fortunately allow for will be derived and compared to
available data. We   mainly discuss   two evolutionary 
measures which are natural in our approach: 
the life-time distribution and the rate of extinction.
From the latter, the distribution of extinction event sizes, 
$G(x)$ can be approximately derived as sketched in the introduction. 
 
\subsection{Record driven dynamics}
 
One might describe a living organism abstractly
in terms of its genome - a string of data
coding for a certain  functionality. From this point of view, each
individual is a point in an extremely large space or
`fitness landscape' \cite{Wright82}, where
genetically  similar individuals  appear as
clusters. Species and higher taxa can be viewed
as clusters at different levels of resolution. 
On appropriate time scales,
the typical genotype, which coarsely describes the genetic pool
(cluster) of a species, moves from
one    metastable configuration 
to another, due to the influence of mutations\cite{Weisbuch92}.

There are  at least two conceptually distinct ways
of producing  punctuated behavior.   One way is based on
an `energy' picture, where the fitness landscape is assumed to have
strong barriers, which  the population has to cross to get to another
fitness peak\cite{Bak93}. This picture introduces long times of stasis, where
the typical genotype of a species does not change.
Another possibility is to emphasize   
 entropic barriers. In this case,    the long time scales are   due not to
fitness barriers, but rather to the simple fact that another fitness
peak may be hard to find, given the enormous number of possible
mutations. More specifically, 
one can expect   a diffusion like   behavior  in genome space along
directions   which are selectively neutral \cite{Kimura83},
while, under stable or metastable conditions,
 mutations in other directions   must be mainly rejected.
Neutral mutations are instrumental in maintaining  population
 diversity and in creating paths  between distant local fitness maxima. 
 In a simple golf metaphor, the energy picture corresponds
to getting the ball over a hill; you have to push the ball against
gravity to get it over to the other side. The entropy picture, on the
other hand, corresponds to getting the ball into the hole, which may
be difficult, even on a flat green. 

We  now consider an extremely simplified picture of   
  evolutionary dynamics:    a single
species (with a population sufficiently large to avoid extinctions
by size fluctuations and other random factors)
evolving  under constant external physical conditions.
The model  will be used as a stepping stone and a basic ingredient 
for the more complicated case involving  co-evolutionary mechanisms.
The main idea is to map evolutionary dynamics into a 
search for   fitness records,   which then  
 correspond  to   evolutionary events.
 This mechanism does not strongly depend on whether  entropic
or energetic mechanism  dominate the population dynamics. 
It  leads to punctuated equilibrium,   because, as  
new records are     harder and harder to find, 
   the system tends to  stay    put in the same
   state for  longer and longer times.  
The idea of an optimization driven dynamics which becomes 
progressively slower has   appeared before. In addition to  
the remark of Raup and Sepkowski already cited, we note that 
Kaufmann\cite{Kauffman95} has explored the idea in the context of 
`long jump ' dynamics on NK landscapes, also    speculating on
its applicability to technological   and  biological evolution.  
  
The exploration of genome space performed by a population of individuals
is akin  to a random walk, with a generation being the unit of time,
and with mutations corresponding to steps taken in different random directions.
If   fitness values change smoothly - or not at all - 
from one site of the landscape to a neighboring site,
there will be connected subsets of genome space within which
fitness values are highly  correlated.
We assume that these correlated volumes  can be visited within
a certain characteristic time scale, choose this   scale as the unit of time
and coarse grain  each correlated volume into one  point.
The coarse graining
leaves us with a rugged fitness landscape, where each
move  leads to a new  fitness value  unrelated
to the  previous one.
Species sit on a local fitness maximum in this landscape: More accurately,
the individuals of the species occupy
a   correlated volume around a fitness peak.

In order to derive a dynamical rule for  
coarse level evolution, let us  return to
individual mutations. With a certain probability,
the rare   mutations
which  give their  bearers a selective advantage
over all other individuals   will establish themselves
throughout the population. 
 Although this process   may   spontaneously regress,
we   assume for simplicity that it   happens  with probability
one and within one (or few) of our coarse grained time step(s).
Our `optimistic' model  makes  each
species the keeper  of the `best'
genome found `so far' in the evolutionary
search   within a  neighborhood of the landscape.

As the rugged  fitness landscape is very high dimensional,
we can neglect the unlikely event of a  random walker   retracing
its steps. In this case, the fitness values successively generated
in  a sequence of  mutations leading from one region to the
other in the landscape     effectively constitute a stream of
independent random numbers.
Furthermore, according to our above  assumption,  the system
evolves if and only if a fitness record is encountered, in which case
the gene pool moves from a local fitness maximum to the new higher maximum.
We call this form of evolutionary dynamics for record driven dynamics.

The evolution  dynamics  for one  species in a fixed environment
and the statistics of records     in   a  series of random numbers
independently drawn from an identical  distribution are -- in our model --
mathematically equivalent.
As it will become clear,   the  statistics  of records  is
largely independent of the choice of   distribution generating 
the fitness values. The size of the jump   also being unimportant, we have a
model with  {\em no}  fitting parameters, aside from the
time unit.

The   highly  idealized one-species system in a constant physical
environment is clearly
far removed from   natural conditions. It has
however been studied  in the laboratory with microbial
cultures of {\em E. Coli}  by Lenski et al.\cite{Lenski94}, 
whose  experiments  provide us  with empirical data
to which   our predictions can be compared.  

\subsection{ The statistics of records}

A great advantage of the simple record model is that it allows for an
exact mathematical description \cite{Sibani93}.
Consider  a sequence of independent random numbers
drawn   from the same distribution at times $1, 2, 3, \ldots, t$.
We exclude distributions supported on a finite set,
because    they  eventually   produce
a record which cannot be beaten.
The first number drawn  is by definition a record.
Subsequent trials lead to a
record if their outcome is larger than   the previous record.

We seek the probability $P_n(t)$
of finding precisely $n$ records in $t$ successive trials, where
$1 \leq n \leq t$. In the derivation we need the auxiliary
function $P_{(1,m_1,\ldots m_{k-1})}(t)$, which is the joint probability
that $k$  records happen     at times $1 < m_1,\ldots < m_{k-1}$,
with $ m_{k-1} \leq t$.
$P_1(t)$ is simply the probability that the first outcome
be  largest among $t$. As each outcome has, by symmetry,
the same  probability of being the largest, it follows that
$P_1(t) = 1/t$.
In order to  obtain two
records at times  $1$ and $m$,   the largest of
the first $m-1$ random numbers must be drawn   at the
very first trial. This
happens with probability $1/(m-1)$. Secondly,
the $m$'th outcome must be the largest among $t$. This
happens  with probability $1/t$, independently   of  
the position of the largest outcome
in  the first $m-1$ trials. Accordingly,
\begin{equation}
P_{(1,m)}(t) =   \frac{1}{  (m-1) t}
\label{P_1m}
\end{equation}
Summing the above over all possible values of $m$ we obtain
\begin{equation}
P_2(t) =  \sum_{m=2}^t \frac{1}{ (m-1) t} \approx \ln(t)/t.
\label{P_2}
\end{equation}
In the more general case of   $n$ events, we similarly obtain
\begin{equation}
P_{(1,m_1\ldots m_{n-1})}(t) = \frac{1}{  \prod_{i=1}^{n-1}(m_i-1) t}
\label{P_km}.
\end{equation}
We now take $q_i = m_i-1$ and sum over all possible values of the
$q_i$'s. This leads to
\begin{equation}
P_n(t) =
   \sum_{q_1=1}^{t-n+1} \frac{1}{q_1} \ldots
  \sum_{q_{n-1} =q_{n-2} +1}^{t-1} \frac{1}{q_{n-1}} \frac{1}{t}.
\label{P_k}
\end{equation}
An approximate  closed form expression can   be obtained  by
replacing  the sums by integrals, which is reasonable
for $t>>n>>1$. The integrals can then  be evaluated
finally  yielding
\begin{equation}
P_n(t) =  \frac{(\ln t)^{n-1}}{(n-1)!} \frac{1}{t}.
\label{log-poisson}
\end{equation}
Interestingly,  Eq.  \ref{log-poisson} is a Poisson
distribution, but with $\ln t$ in place of the more
usual $t$.

Let $\overline{n(t)}$ and $\sigma_n^2(t)$ be the average and
variance of the number of   events in time $t$.
As an immediate consequence of Eq.  \ref{log-poisson} we note that
\begin{equation}
\overline{n(t)} = \sigma_n^2(t) = \ln t.
\end{equation}
If one assumes that each evolutionary event carries a fixed amount
of `improvement', in a sense made precise later,
measured averaged fitness and fitness variance will have the same
type of time dependence. We also note that the `current',
i.e. the average number of events per unit of time decays as
\begin{equation}
\frac{d\overline{n}}{dt} = \frac{1}{t}.
\label{oneovert}
\end{equation}

Another    consequence  of Eq.  \ref{log-poisson} is the following:
Let $t_1 =1 < t_2 <  \ldots < t_k < \ldots $ be the times at which
the record breaking events occur, and let
$\tau_1=\ln t_1=0 < \tau_2  \ldots < \tau_k=\ln t_k < \dots  $ be the
corresponding
natural  logarithms.
The stochastic variables $\Delta_k = \tau_{k+1}-\tau_k=\ln (t_{k+1}/t_k)$
are independent and
identically  distributed. Their common distribution is an exponential with
unit average.
By writing: $ \tau_k = \Delta_{k-1} + \Delta_{k-2} + \ldots \Delta_1 $
we have that $(\tau_{k+1} - k)/\surd k$ approaches a
standard gaussian distribution for large $k$. Hence, the waiting time $t_k$
for the
$k$'th  event is approximately log-normal, and the average of its  logarithm
grows linearly in $k$. By Jensen's inequality \cite{Rudin66}   we also find
\begin{equation}
\ln ( \overline{t_k} ) \geq \overline{ \ln(t_k)} = k
\label{exp_bound}
\end{equation}
We see that the average waiting time from the $k$' th to the  $k+1$ 'st
event grows at least exponentially in $k$.

Consider finally  the case in which a system is made up
of $p$ several parts
which act independently of each other.
This could for example happen  in  a very large population
which splits into
subpopulations occupying  geographically separated areas.
The total number of record events within  a certain period
of time is the sum of  the events  affecting   each
subpopulation. This sum
remains Poisson distributed, with average  $p\log t$ rather than
$\log t$ and our   conclusions remain therefore 
 unchanged up to a trivial scale factor.

\subsection {Evolution revisited} 
We first note that the record dynamics gives rise to an event rate which,
according to Eq.  \ref{oneovert}, decays as a power law. In the
most simple scheme, where a fraction of the evolutionary   events
leads to extinctions,
this would indeed imply a power-law decay of the extinction rate. However, the
exponent would be $\gamma=1$, and not $\gamma \approx 0.4$ as suggested by the fossil
data. It would also be surprising if the exponents coincided, since no
ecological or co-evolutionary  interactions among  species   have yet been considered.
In the next chapter we shall see how such interactions may change the
extinction rate.

The logarithmic growth  of fitness in a population evolving under constant
external conditions has been observed in experiments by
Lenski and Travisano\cite{Lenski94}. These authors  considered  12
populations of {\em E. Coli},
which were all generated by identical  (cloned) ancestors,
and were allowed to grow in identical but physically separated environments.
They investigated  the change in mean cell volume and fitness, the
latter defined as the Malthusian rate of population increase.
The data depicted in    Fig.~2 are taken from  Ref.~\cite{Lenski94}.
They  show the   time  dependence of
the average fitness in a single population of {\em E. Coli}, relative to its
original ancestor.    We have redrawn these data on  a semilogarithmic scale,
to demonstrate  the agreement  with the  logarithmic growth  predicted
by our  record driven dynamics.
Lenski and Travisano also
 studied the variance of the mean fitness and cell volume across
different populations,
and found them to be growing. Our model   predicts a logarithmic growth
for the fitness variance as well as for any other quantities which changes
- on average - by a constant amount upon an evolutionary improvement.

\begin{figure}
 \centerline{\psfig{figure=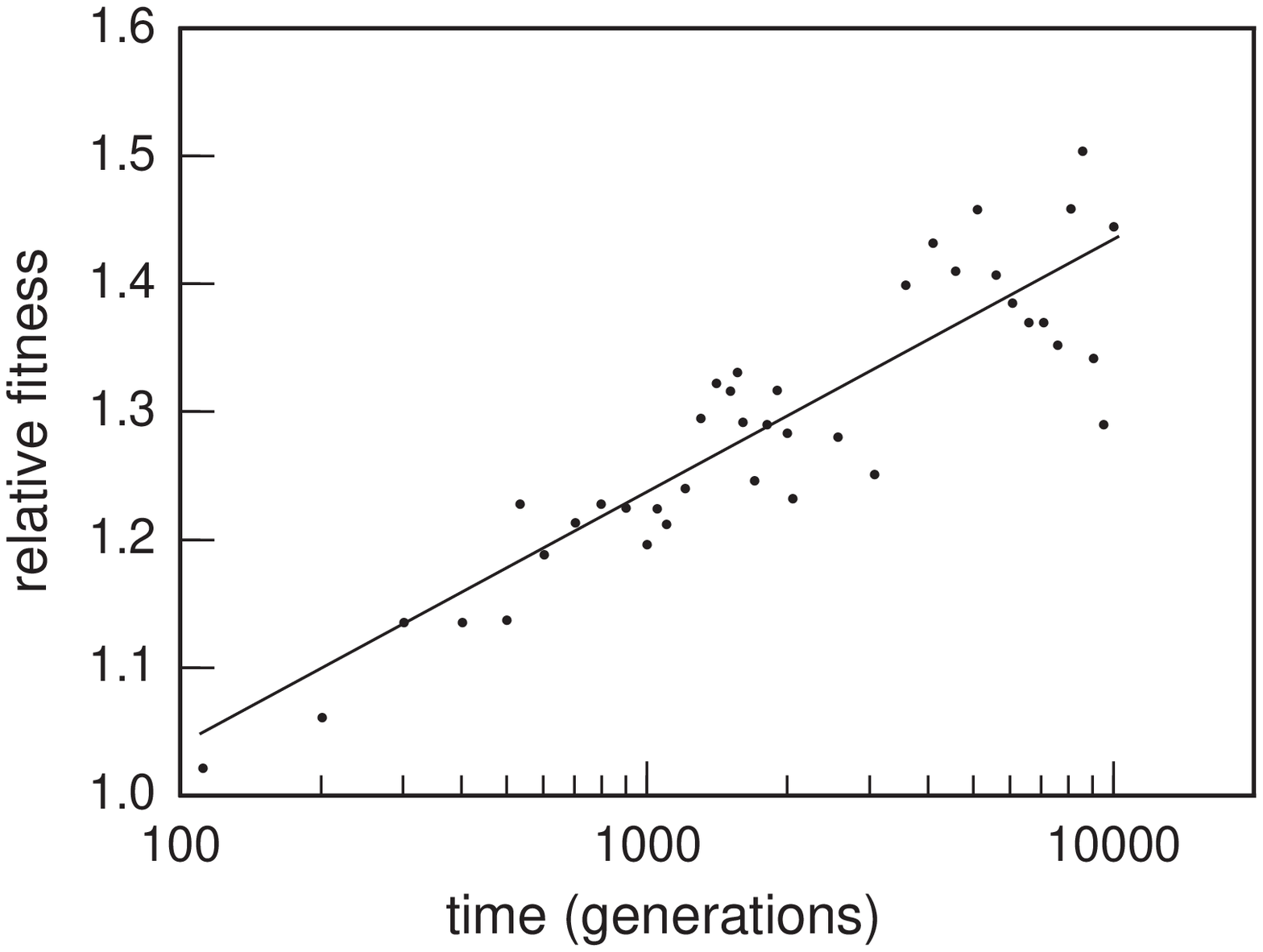,height=10cm,width=14cm}}
\begin{caption} {} 
\begin{scriptsize} 
The   mean fitness in one population of {\it E. Coli}
grown in a glucose-limited environment, is shown in a semilogarithmic
plot as a function of the number of generations. The original data
 and a detailed
description of the controlled evolution 
experiment can be found  in Ref.~\cite{Lenski94}.
Note that the data approximately follow the logarithmic fitness growth curve
discussed in connection with the reset model.
\end{scriptsize}    
\end{caption} 
\end{figure} 
\section{The reset model}
The `reset' model 
includes  the idea that macroevolutionary events are triggered by
fitness records, and complements  it with  the possibility of extinctions
through interspecies interactions in an ecological space.
The   model deals
with a number of highly idealized species
evolving and interacting with
one other within an `ecology'.  The whole system
is conveniently depicted as
a graph, in which  nodes correspond to  niches and
edges represent interactions among the (neighbor)
species which occupy
them. Model  species are born, evolve and eventually die.
In most of the treatment  below the (physical) external
conditions are assumed constant. However,
we additionally   consider a `meteor' variant of the model
where    the effects of changes
in  external conditions are modeled as
random  killings (i.e. unrelated to    fitness )
of a fraction of the systems'  species.

The  evolution part of the  reset model is basically
the   record dynamics   just
described. Between    evolutionary changes,
no  extinctions are allowed
and the system remains quiescent.
When a species evolves, it is assumed to
modify  the fitness landscape  of its
ecological neighbors, thus creating the
possibility of    extinctions. Consider
a species
evolving  from fitness $f$ to $f' > f$. By definition
of the model update rule,
those among  its neighbors  having fitness
less than $f'$ are declared extinct  and removed from the system.
In the next time step,   the empty niche is
`reset', i.e. it becomes occupied by a freshly
initialized species.
This defines the reset model. Note that the reset   
rule depends on the rank ordering of the agents,   not on the
specific form of   the fitness. 
 We have implemented versions of the model
with slight   differences   in the details
of the removal and replacement of the species.
However, we find the results to be robust to these changes.

In the simulations discussed below, the ecology  is modeled as  a regular
lattice. Previously, we have investigated both
one and  two dimensional systems \cite{Sibani95},
obtaining very similar results, so we here  confine ourselves
to a 2D grid with  unit
lattice constant. In this grid, the neighborhood
of the point  $x_0,y_0$ is defined as the set of integers:
${\cal N}_0 = \{ (x,y) \}$ such that
$ |x-x_0| \leq CR$  and $|y-y_0| \leq CR $.
The only tunable model parameter is
the Coupling Range, $CR$:
If $CR=0$ our system reduces to a set of uncoupled
agents, each searching for records in its own fitness landscape,
while if  $CR =\infty$ all agents are coupled to each other.

The reset dynamics could  be simulated time step by time step,
every time assigning for each agent a random trial fitness value,
drawn from a given distribution. However, in this implementation
nothing  happens until the trial
fitness value of a given agent exceeds the current  
value, a process   which is computationally inefficient.
 Instead, we shall make use of some
analytical results described below, allowing us
to skip the inactive periods. 

\subsection{Fitness records and waiting times}
Our   simple  choice  is to draw all
fitness values from a uniform distribution in the unit
interval. As the first value drawn
is by definition the  first record $f_1$,
this quantity  is also   uniformly distributed
in the unit interval. The $k+1$'th record  is required to exceed
its predecessor  and is therefore    uniformly
distributed in  $(f_k ,1)$. The conditional probability density
$p_f(x|k)$ that
$f_{k+1}=x$ after $k$ records is given by :
\begin{equation}
p_f(x|k) = \frac{-\ln(1-x)^{k}}{k!} \; \; \; 0 \leq x<1.
\label{dist1}
\end{equation}
The formula   is trivially true    for $k=0$. For $k > 0$ it
suffices to note that
Eq.  \ref{dist1} solves the  recursion relation
$p_f(x|k) = \int_0^x p_f(x'|k-1) (1-x')^{-1} dx'$, where
the factor $(1-x')^{-1}$ provides the proper normalization
of a uniform density  in the interval $(x',1)$.

For later convenience  we use the
related fitness  measure
\mbox{ $z = -\ln (1 -f)$}. The probability density  $g_z(x|k)$,
 that $z=x$ after $k$ records is found by
a change of
variables in  Eq.  \ref{dist1}
yielding
\begin{equation}
g_z(x|k) = \frac{x^{k} \exp (-x) }{k!} \; \; \; 0 \leq x<\infty.
\label{dist2}
\end{equation}
The form of  $g_z(x|k)$ indicates that
$z|k$  arises as the  sum of  $k $ independent
and identically distributed variables
$\Delta_1, \ldots , \Delta_k$, each  describing
the fitness increment in   one   evolutionary step.
Their common distribution is an exponential of unit  average:
\begin{equation}
p_{\Delta}(x) = \exp (-x) \; \; \; 0 \leq x< \infty.
\label{incr}
\end{equation}
It immediately
follows from Eq.  \ref{dist2} that the  average of
$z$  grows linearly with the number of events and hence
logarithmically in time.

Let us finally consider the distribution of the
waiting times for the next record to happen
given that the system   has    fitness $f$,
with $0\leq f <1$.
We assume  that each trial takes one time unit.  With   probability $f$
an attempt does {\em not} lead to a   record.
The attempts being independent,
the conditional  probability  for the  waiting time to
next record being   $\delta$ is:
\begin{equation}
p_w(\delta|f) = f^{\delta-1} (1-f) \; \; \; \delta \in \{ 1, 2 \ldots \}
\label{waiting}
\end{equation}
 
\subsection{The reset algorithm}
We are now ready to define a computationally  efficient algorithm for the
reset model.
Although  two equivalent  fitness measures are in use:
$f_i$ and $z_i$, with $z_i = -\log (1-f_i)$
we only explicitly mention the latter in the following description.

At any fixed point of time  an  agent $i$ has two attributes:
the  fitness $z_i$, and the step time $t_i$ at which its next evolutionary
step will be taken -- unless   the agent is  killed at an earlier time
by evolution in a neighbor site.
Initially, the time is $t=0$  and
all agents have fitness $z_i =0$,  which means that there are no
`living' species in the system. The step-times $t_i$ for the next
move are all initially set to the waiting time $\delta_i=1$, in
accordance with Eq.  \ref{waiting}.
The  core of the algorithm now  iterates the following steps:
\begin{enumerate}

\item Move time to $t = \min_i \{ t_i \}$

\item Pick the agent(s) $a_k$ with  $t_k = t$, and update its (their)
fitness and step-time(s):

	\begin{enumerate}
		\item Generate the fitness change	$\Delta_k$ according to Eq.  \ref{incr}
		\item Update fitness: $z_k  \rightarrow z_k +\Delta_k$
		\item Generate the waiting time $\delta_k$ according to Eq.  \ref{waiting}
		\item Update the step-time: $t_k    \rightarrow t_k +  \delta_k$
	 
	\end{enumerate}

\item Select unfit neighbors and reset them:
%For all   updated agents $a_k$:

	\begin{enumerate}
	 \item A neighbor $a_j$ of an updated agent $a_k$ is selected as unfit if
                $z_j<z_k$ 
 
		\item Reset fitness of unfit agents $a_j$: $z_j \rightarrow 0$
       
		\item Reset step-time of unfit agents: $t_j \rightarrow t +1$
       
	\end{enumerate}

\end{enumerate}

Note that in one single pass an agent can   {\em both} be  updated
in fitness  {\em and}
become tagged as unfit due to the evolution of
one of its neighbors.
This version of the reset  rule    is  insensitive
to the sequence in which these  two   events   take place,
which has some importance  at short times, when the activity is high.
Later,   as  evolutionary events  thin out, it becomes
increasingly unlikely that two neighbor sites would evolve in the same pass.
Also note that the algorithm  
 skips the increasingly long intervals of time where
the system remains quiescent. This device  makes it possible  to run
simulations spanning a  large number of time decades.
During the runs we  collect a variety of statistics: e.g. the number of
extinctions,   the life-time of species, with the statistics gathered
during a time window of   selectable width, and the  
 number of improvements that  agents undergo during their life-time.
Other statistical measures are the average fitness in the system as a
function of time and the way in which empty niches are filled as the
system evolves.

\subsection{Analytical properties of the reset model}
In a later section we present analytical results  for  a
{\em mean field  } version of the reset model.
Here we like to   mention a couple of properties
of the full model  which are
easily  derived:
1) punctuation in spite of the absence  of stationary behavior on logarithmic
time scales and
2) certainty of death for every agent.

Regarding 1)  we note that the largest fitness value
in the system $z_{max}$  is an increasing  function of time.
Indeed, when  the   agent `carrying'  this value changes
its state, it  either performs    an evolutionary step,
or it is killed by an evolving  neighbor. In the former case
$z_{max}$ clearly increases. In the latter,
the neighbors'  new fitness value $z'$  must
exceed  the previous   $z_{max}$,
thus becoming  the new $z_{max}$.
As $\overline{z} = N^{-1 } \sum_i z_i > N^{-1 }z_{max} $,
we see that the average fitness value must  increase as well.
This  eliminates the possibility that a system
with finite $N$  will ever reach a stationary state, although
it must be kept in mind that the increase will be extremely
slow - i.e. logarithmic -- and thus hardly perceptible
at sufficiently long times.

Regarding 2), we only need  consider the probability that the
highest ranking agent
in a system  of two agents will eventually die, as the
probability of being killed  clearly  grows with the number of
neighbors.
The killing must happen -- if at all --
for some $n \geq 1$,
between the    $n $'th  and   $n+1$'st record of the `victim'.
Let us   name  these time intervals  `epochs'.
We bound from above the probability $S(n)$ that the
agent survive $n$ epochs and show that $S(n)$  vanishes for
$n \rightarrow \infty $. First we calculate the probability $R(n)$
that the agent be  killed during its  $n$'th epoch, given that
it was alive at the beginning of the epoch.

Let the epoch be fixed and the fitness
$f$ (or $z$) be given. Assume for the moment  that
the agent  waits  exactly $t$ steps before its next improvement.
The probability that he is  meanwhile   overcome by
his neighbor and therefore    killed is
\begin{equation}
R(t,f,n) =  \sum_{l=1}^t f ^{l-1} (1-f ) = 1-f^t.
\label{prob2}
\end{equation}
We  first  average over all possible values of $t$,
with a weight given by
Eq.  \ref{waiting}, to  obtain the  probability  that
an  agent with fitness $f$   is  reset in its $n$'th epoch:
\begin{equation}
R(f,n) =  \frac{1}{ 1+f }.
\label{prob3}
\end{equation}
Averaging Eq.  \ref{prob3} over  $f$ according
to  Eqs.  \ref{dist1} (and Eq.  \ref{dist2} after a convenient change
of variable) yields the probability of being killed  during  epoch $n$:

\begin{eqnarray}
R(n) &= & \int_0^\infty \frac{z^{n-1}\exp (-z)}{(n-1)!(2-\exp (-z))} dz\\
     &=& \sum_{k=1}^\infty \frac{1}{2^k k^n}
\label{prob4}
\end{eqnarray}

The probability of   being
killed  before   the  first
improvement is   $R(1) = \ln 2$.
We also note  that  $R(n) > 1/2$ for all $n$.\
Finally, the probability of surviving
$n$ epochs is
\begin{equation}
S(n) =  \Pi_{l=1}^{n} (1-R(l)) < (1/2)^{n+1},
\label{prob5}
\end{equation}
which vanishes in the limit $n\rightarrow \infty $ as claimed.

\subsection{Simulation results}

In this section we describe the behavior of the
reset model  obtained through numerical simulations.
The data shown in this section represent
a substantial numerical effort - of the order of
one year of continuous calculation on a dedicated
 workstation. One single simulation took a full
half year.
Rather than reusing  and modifying  the Pascal code
used in Ref.~\cite{Sibani95},
completely new   programs
where developed in C  by     one of the authors (M. B.).
This provided increased portability as well as an
independent check.  The current results
concur with our previous simulation  and extend them
in several ways: The new
simulations are considerably  longer, revealing
new features in the data and probe in addition
the   effect of
externally imposed  catastrophes (`meteors').

Throughout the sequel the symbols $\log$ and $\ln$ stand  for
base $10$ and   natural logarithm respectively.
We generally use $\log t$ as abscissa. As an inconvenient
side effect, functions proportional to
the natural logarithm of time appear  in the plots
as  straight lines with slope
$\ln 10 = 2.3$. All data shown stem from simulation of a system of
2500 agents located on a $50 \times 50$ grid.
Plotted fitness values are always the   $z$ version
described right above Eq.  \ref{dist2}.
An agent is defined as being active if it has fitness
larger than zero.

In Fig.~3 we show, as a function of time
a)   the number of active agents;
b)  the average fitness  and   c)  
 the minimum and d) the maximum   fitness  
 value among the active agents. 
As indicated, in each graph data are shown for
$CR=0$, i.e.   no coupling, $CR=2$, and $CR = 4$.
When $CR=0$ all agents are and remain
active, except for the very first  time step. The average fitness
$\overline{z}$ is  proportional  to $\ln t$. The minimum 
and maximum fitness shown in plate c) and d) do
the same. The punctuated behavior of the maximum fitness
in the system is very clear. 
With $CR =2$ the number of active agents first saturates at
 $t \approx  10^5$. The average fitness   goes through
a cross-over at  $t \approx 10^3$, with the slope
changing from slightly above  $1$ to $2.3$, which is the
value  for the uncoupled system.
While the punctuated motion of the maximal fitness qualitatively
is indistinguishable from the $CR=0$ case, it is clear that the
minimal fitness stays close to zero up to $t=10^5$, as also
expected from plate a).
Similar behavior is seen for $CR=4$, except that the cross-over
in average fitness appears  at about $t=10^7$. Saturation
of active agents is hardly seen, and the minimal fitness never increases.
\clearpage
\begin{figure}[tp]
 \centerline{\psfig{figure=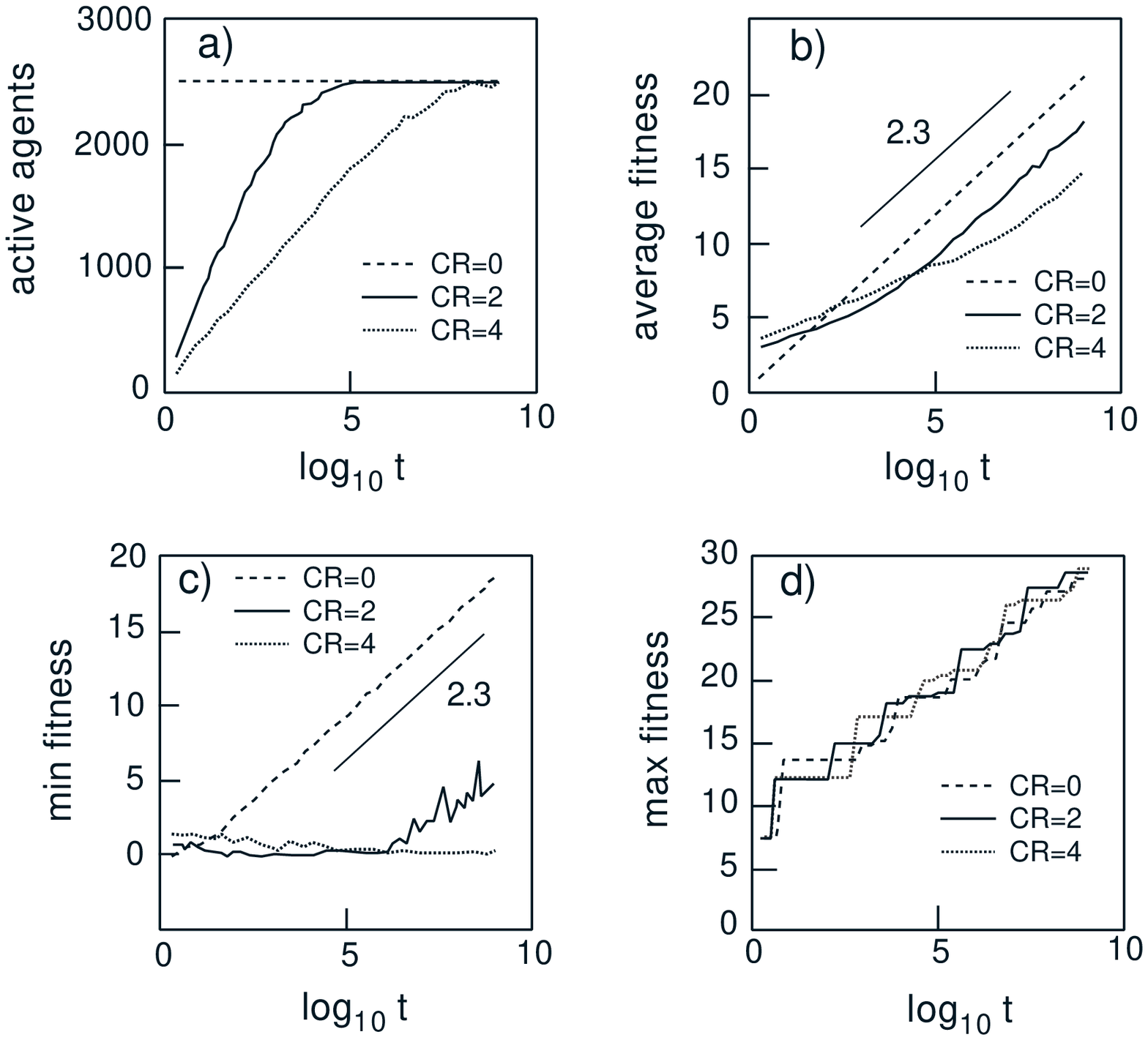,height=12.5cm,width=14cm}}
\begin{caption} {}  
\begin{scriptsize}  
As a function of time we plot:
a) the number of active agents, b) the average fitness,   c)
the   smallest fitness value among active agents and d)
the largest fitness value among active agents.  
As indicated, in each of the subplots, different lines correspond
to different degrees of coupling in the system. 
\end{scriptsize} 
\end{caption}  
\end{figure}

In conclusion, we see that   the systems' behavior with respect to
fitness eventually approaches that  of an uncoupled system,  
with the fitness distribution moving with a velocity proportional
to $\ln t$. 
Figure 4 describes the time  dependence  of the rate of extinction
  for a series of different coupling lengths. (In a stationary system,
   this quantity would remain constant)
We   note that coupling length $CR $ strongly affects the shape of the curves.
For $CR = 1$ the decay is -- after a few decades
 -- a power law with exponent $-1$.
It is clear that for $CR=2$ and $3$,  the data eventually reach the same
asymptotic behavior. Based also on the previous results,
we would guess that -- independently of $CR$ --
all curves reach  the same asymptotic  behavior. Note that, according
to Eq.  \ref{oneovert}, $-1$
is the exponent characterizing the uncoupled
record dynamics, if one relates extinctions and evolution by a
 proportionality factor.
Considering the extremely large number of decades involved, it is  however
doubtful  that the asymptotic behavior is the relevant one.
\clearpage
\begin{figure}[t]
\centerline{\psfig{figure=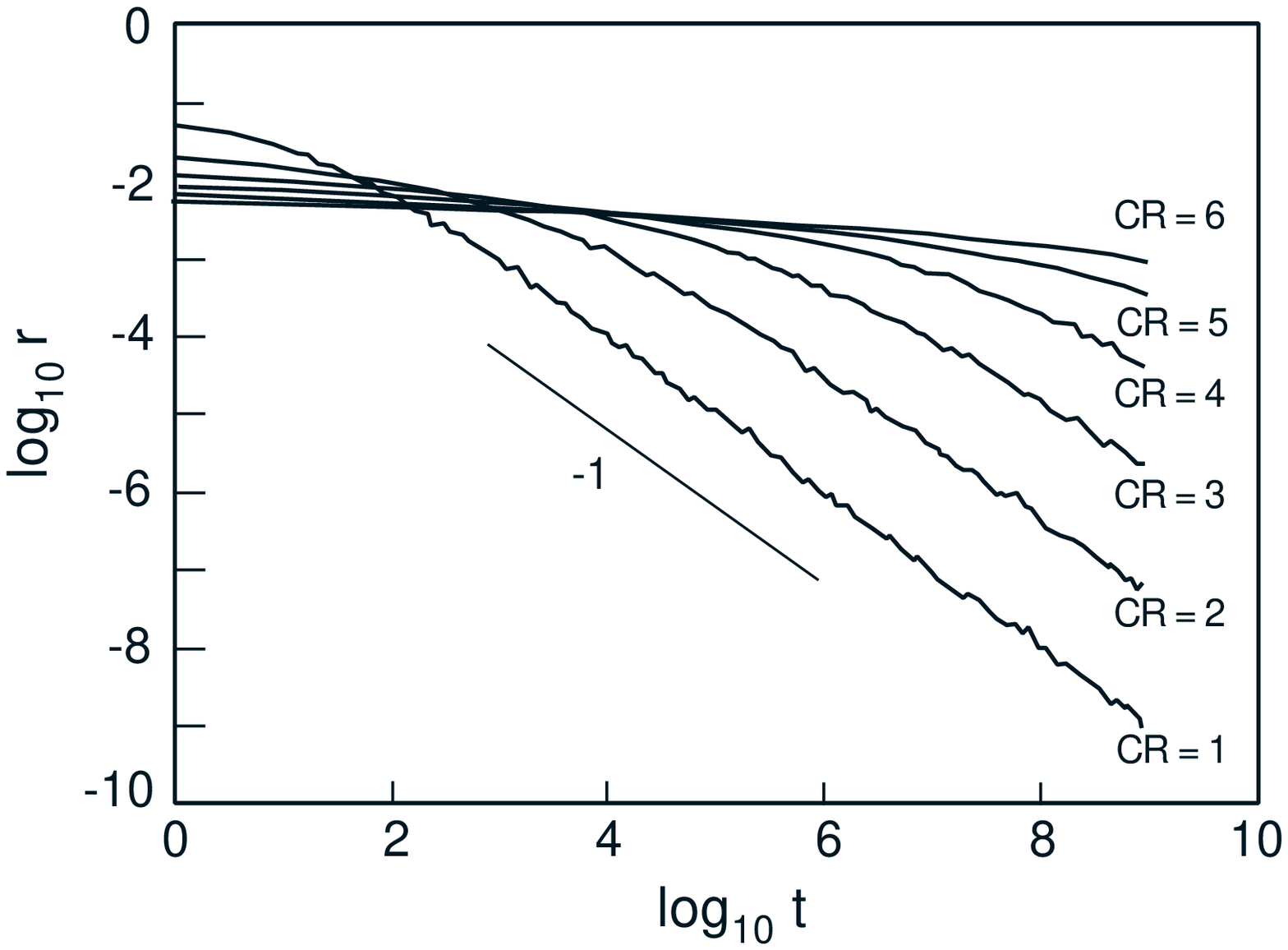,height=10cm,width=14cm}}
\begin{caption} {}
\begin{small}  
 The logarithm of the rate of extinction  ( calculated as the fraction of extinctions
 taking place in each time bin )  is shown  versus the natural logarithm 
 of time for a series of different coupling
 lengths. For extremely long times the plots suggest that all the  systems
 would eventually approach an algebraic decay with exponent $-1$, which corresponds
  to the behavior of the weakly coupled system. However, for many intermediate time
  scales  the coupling length strongly affects the shape
 of the curves, with the effective decay exponent becoming progressively smaller as
 the coupling strength increases.  
\end{small}
\end{caption}  
\end{figure}
\noindent Over many decades (more than the empirical data can offer), the
rate of extinction decays with an effective exponent which
is numerically much smaller than $1$, and consistent with fossil data.
We return to this issue in the next chapter in connection with the analysis
of a continuum model \cite{Sibani97}.

Plates a) to f) in Fig.~5 each depict   life span distributions
in systems with $CR$ ranging from $1$ to $6$. Each plate  contains
nine different  data sets, as
most clearly seen in  plate a). Each of these   is the distribution of
the life spans  collected in a time window
limited by   system age $10^0$ -- the origin of
time -- and $ 10^i$, with $i=1, 2, 3, \ldots , 9$.
Thus, the topmost graph is the life span  distribution of all the
species which died during
the entire simulation. In the lowest graph only   species which  died
during the
first decade of simulation are counted. From plate d) on, i.e.  for a
coupling range $CR \leq 4$,
all distributions are basically power-laws with an exponent of $-2$.
This is again in good agreement with fossil data (Fig.~1d). 
  At short times, a regime with a numerically lower exponent is
visible, but only  for small values of $CR$.
\clearpage 
\begin{figure}[hph]
\centerline{\psfig{figure=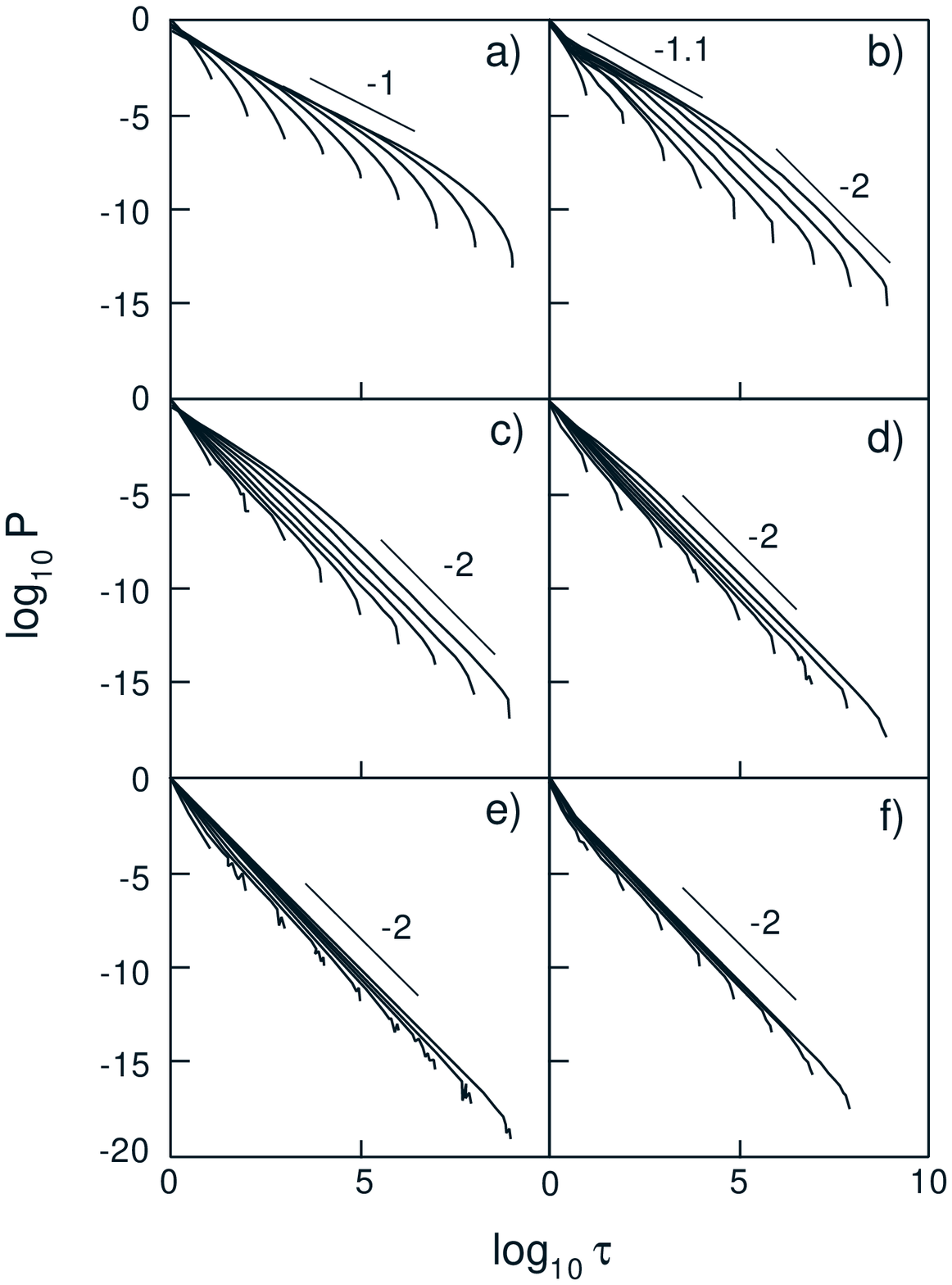,height=19cm,width=14cm}}
\begin{caption} {}
\begin{scriptsize}  
This figure  shows  the   logarithm of the  life span 
distributions of the model species versus the   logarithm of
time. We investigate the effect of varying 
 the  coupling length, which  increases from  plate a) with
 $CR=1$ to plate f), with  $CR = 6$. 
In each case, we show the effect of different ways of
collecting the statistics, an effect   most clearly seen in plate a).
In each plate we have nine different data sets: the topmost curve is 
the distribution of life spans collected in a time window stretching from
$10^0$, the origin, to $10^i$, with $i=1,2,3 \ldots 9$.
Note that from plate d) on, all distributions are basically power-laws 
with exponent $-2$. 
\end{scriptsize}
 \end{caption}
\end{figure} 
\clearpage 
\begin{figure}[hph]
\centerline{\psfig{figure=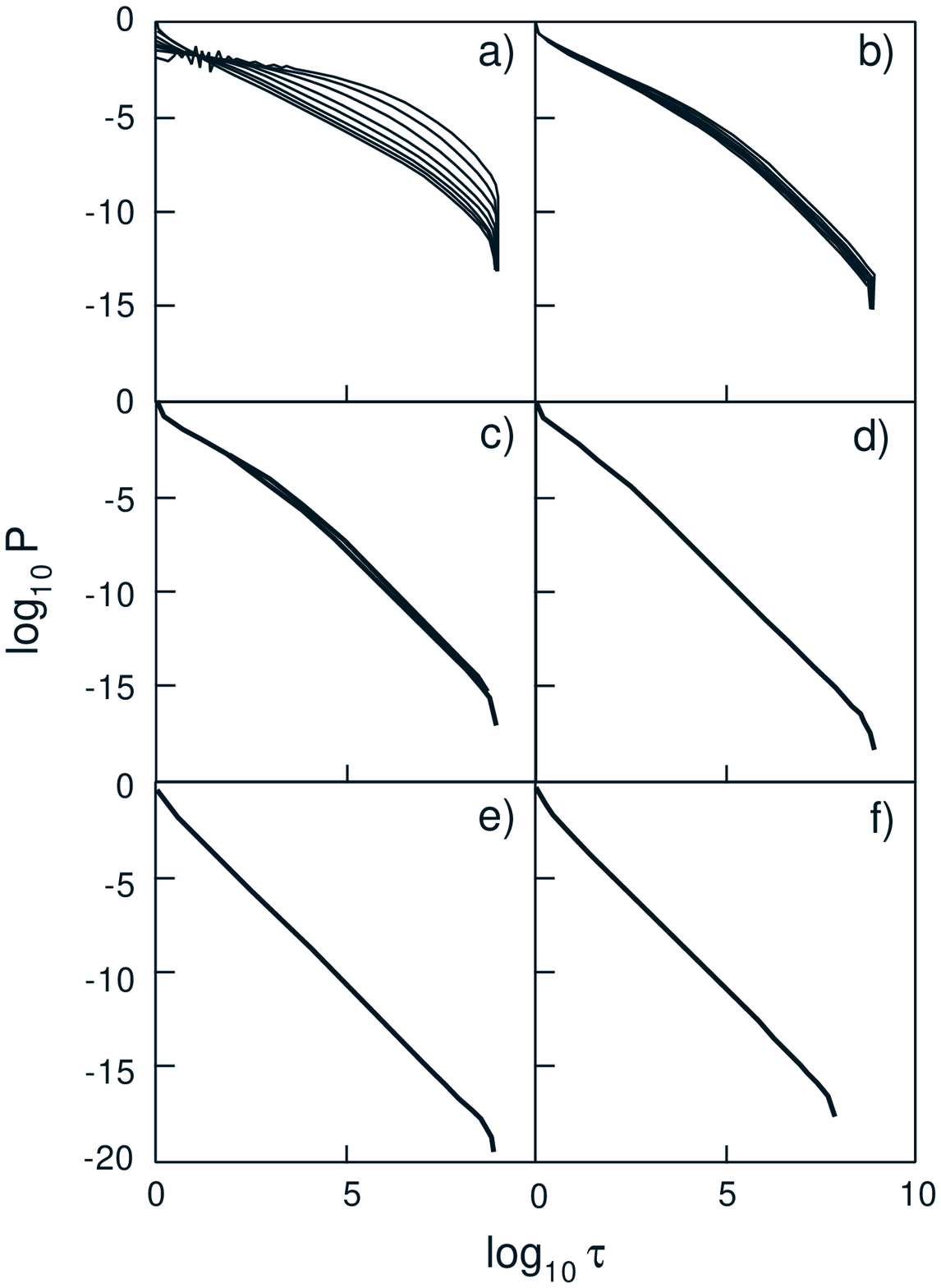,height=19cm,width=14cm}}
\begin{caption} {}
\begin{scriptsize}  
This figure  is similar to Fig.~5. It  shows life spans distribution
 with  the  coupling length increasing  from  plate a), 
with  $CR=1$ to plate f), with $CR = 6$. 
We again  have nine different data sets in each plate.
Numbering them from the topmost down, the $i$'th data set 
is collected from time $10^i$ to $10^9$, which is the  
duration of the whole simulation. Again, we see that the
$t^{-2}$ behavior robustly appears. 
\end{scriptsize}
\end{caption} 
\end{figure} 
\clearpage
\noindent Figure 6 also  depicts life-span distributions. It is
organized in the same way as Fig.~5, with   plates  a) to
f) describing $CR$ values   from $1$ to $6$. The   definition
of the windows in system age during which the statistics is collected
is however different from and complementary to that of  Fig.~5.
Referring for convenience to plate a), where the data sets  are
more easily  distinguishable,
the top most graph
describes the life spans of agents which died between system
age $10^8$ and $10^9$. The next graph pertains to
those dying from $10^7$ to $10^9$ and so forth,
down to the bottom one, which is
identical to the top curve of the corresponding plate in Fig. 5,
and which incorporates events
happening during the entire run,  from $10^0$ to $10^9$.
Again, we see that the power law behavior with exponent $-2$
characterizes data with $CR \geq 4$.
In the  mean field  model described below \cite{Sibani97},
the life-span distribution of a
cohort -- that is a set of agent born at the same time --
is always a power law with exponent $-2$.
Averaging over the  time at which agents are born
does not change this exponent,
provided that the rate of extinction does not change
`too drastically' during the life-time of the
agents. This concurs with the numerical
results shown here, and  further  supports
our  explanation of the  scaling form 
of the  life span distribution obtained 
from  the fossil record. 
\begin{figure}[t]
\centerline{\psfig{figure=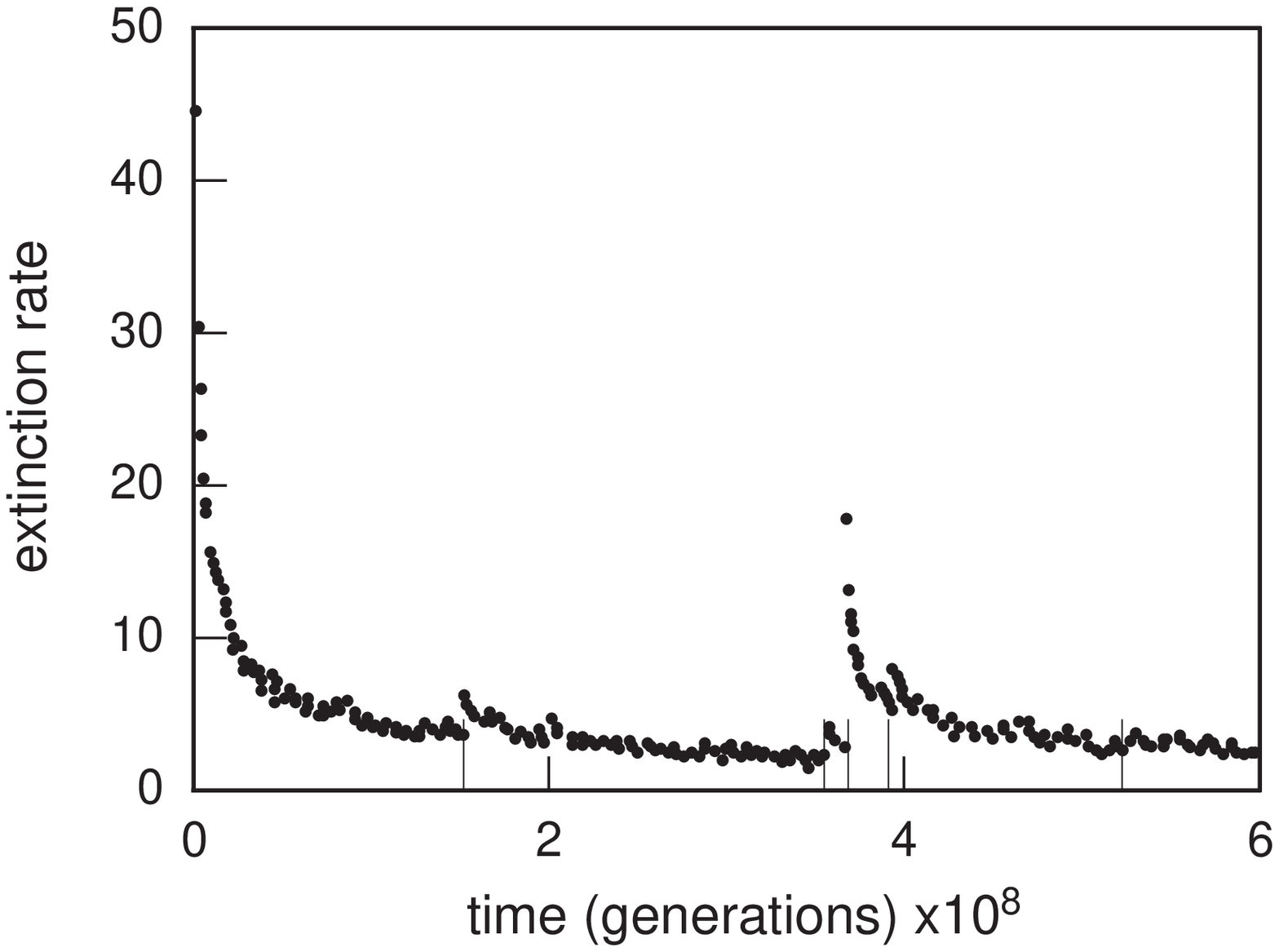,height=10cm,width=14cm}}
\begin{caption} {}
\begin{scriptsize} 
We  show the rate of extinction in the model 
as a function of time in a system
subject to five external five catastrophes  in which 
$12, 14, 52, 12 $ and $11$ \% of the extanct species are
killed at random (independently of their fitness)
and replaced by others. The onset of the mass extinctions is 
marked by thin vertical line segments.
Note the strong rebound effect after the mass extinctions, which
is superimposed on the decaying background rate.   
 \end{scriptsize}  
\end{caption}
\end{figure}

In order to describe  the effect of external `catastrophes'
we simulated  a modified version of the reset
model, where -- on top of the usual reset dynamics -- 
a randomly chosen  set of agents is
destroyed and  replaced at
certain times $t_1 \ldots t_k \ldots$. This replacement
 differs from the usual
extinction  rule in that it is indifferent to  evolution,
i.e. species with high and low fitness are killed  with
equal probability. Furthermore all the affected species
are removed  in a single  time step, as in a mass extinction.
 \begin{figure}[h]
\centerline{\psfig{figure=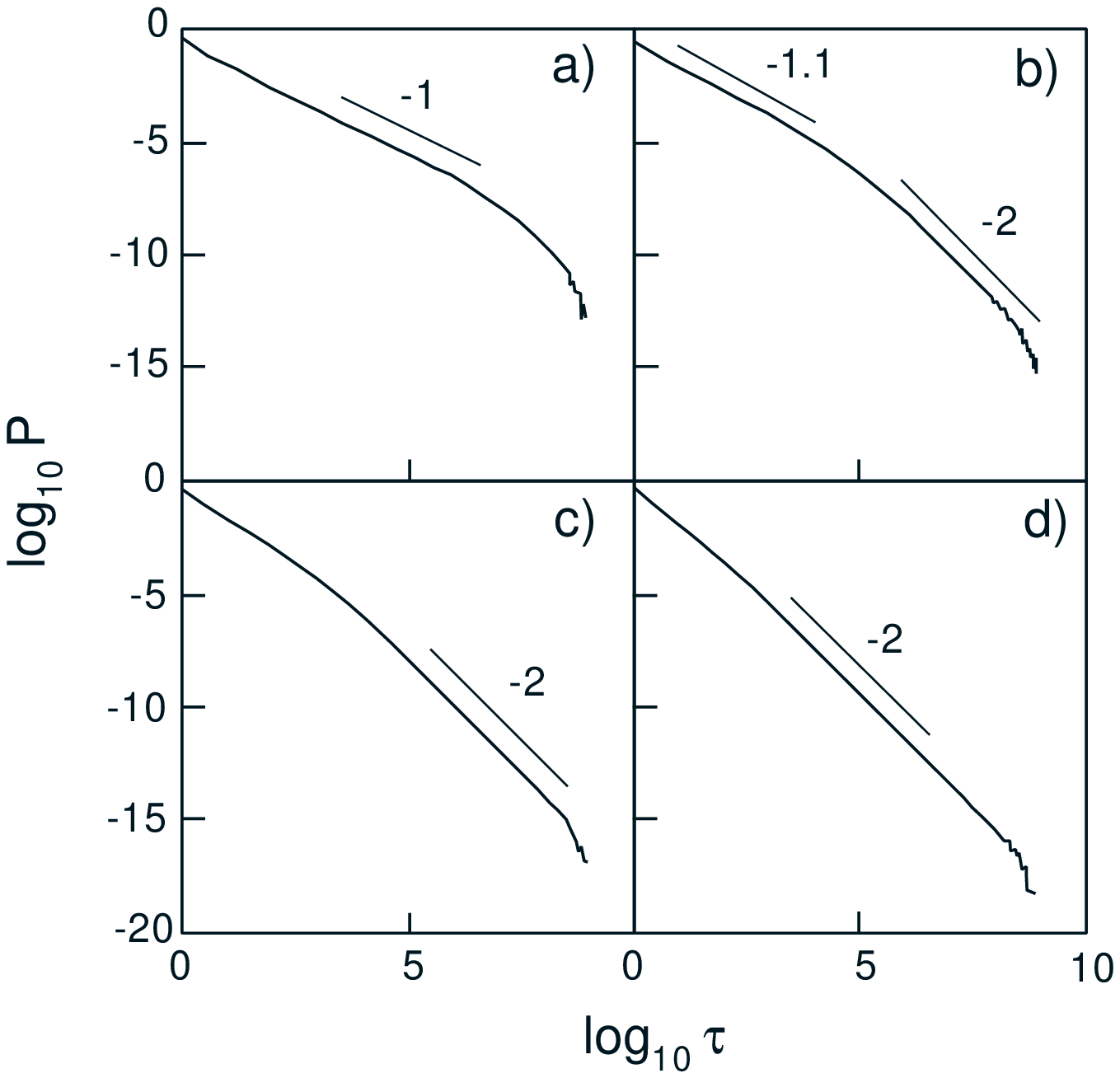,height=14cm,width=14cm}}
\begin{caption} {}
 \begin{scriptsize}
We again show life-time distributions for four different 
coupling lengths ranging from $CR=1$ in plate a) to $CR=4$ in
plate d). Even though these plots are almost indistinguishable 
from those of the previous two figures, the dynamics has been
changed considerably by introducing five `external' random catastrophes,
where $12, 14, 52, 12 $ and $11$ \% of the extanct species
are chosen at random and removed. The corresponding times were
$152.2, 358.2, 367.1, 394.0$ and $528.4$ millions of steps.
These figures show that the $t^{-2}$ behavior of the life time
distribution is completely uneffected by even large external events.  
\end{scriptsize}
\end{caption} 
\end{figure} 
We chose to introduce catastrophes  roughly similar in spacing
and size to the five big mass extinctions. In millions of
generations  the parameters $t_1 \ldots
t_k$ were $152.2$, $358.2$,  $367.1$, $394.0$, and $528.4$. The severity of
the events measured as a  percent of the total extant species
which were killed
was $12, 14, 52, 12 $, and $11$.
The  massive events  have  a clear effect on the rate 
of extinction:    randomizing   the system 
 makes  the pace of evolution higher.   Bursts 
 of evolutionary activity  after each big
event are  shown in  Fig.~7, which depicts the rate of extinction
in the relevant time window for a system with $CR=4$.
 The time is expressed in hundreds of millions of generation
 since the start of the simulation. Thin vertical line segments
 mark the onset of  catastrophes. 

 The behavior of the 
life span  distributions is  shown in the
four plates  of Fig.~8, where the coupling range varies  from
a) $CR=1$ to d) $CR=4$. These curves are almost indistinguishable
from their counterparts  without catastrophes.
The reason is simple: In the model, as well as in reality, the overwhelming
majority of  extinctions   happens outside the big events. 
     
\section{The mean field model}
Some analytical insight in the  behavior of the reset model can
be gained by studying a mean  field version, of course 
 at the price of 
 introducing  further simplifications. 
The   extended model has one tunable parameter,
the coupling length, which gauges the strength of interspecies
interactions. If this parameter is large  enough, i.e. four
or above,  its precise value has little  influence
on the behavior of the reset model. 
 It seems therefore natural 
 to investigate  the extreme  limit in
which all species are coupled together. In this limit,
it is possible to formulate
a mean field theory as  a  partial differential
equation for the  distribution of fitness values, $P(z,t) dz$.

The material in this section  mainly  follows a brief account
previously published in \cite{Sibani97}.
In addition  we consider a  model variant 
with a different form of  the  killing term.  
This variant has predictions in disagreement 
with the empirical data, but it is nevertheless included  
to illustrate, within this modeling context,
 the negative effect of  choosing a symmetric
type of interaction.

\subsection{Generalities}
The partial differential equations
 defining the time evolution of $P$ are designed
 to mimic the behavior of the reset model,
where -- as we have  shown -- a suitably defiend average fitness
grows logarithmically in time. The additional fact that the
 distribution does not spread appreciably\cite{Schmidt95} during
 the course of evolution 
suggests that one should consider a
{\em first order} transport partial differential equations
(PDE).

In the  simpler limiting  case where  no extinctions take place,
we just have hill climbing in a random fitness landscape.
As remarked right after Eq. \ref{incr}, the average fitness
then increases logarithmically in time, whence the average
velocity fulfills $v = 1/t = \exp(-z)$. This is generally true
if each improvement is on average of the same size, a size 
which we   for convenience assume equal to one.
 
With no interactions, each agent moves in fitness deterministically
and independently of the others. The initial fitness distribution
is therefore simply  shifted  in (log) time, as expressed by the
following partial differential equation:
$\partial  P(z,t) / \partial t + \partial(v(z) P(z,t)) / \partial z = 0.$
We    introduce  interactions via   a  term
$ - g P(z,t) K(P(z,t))$, where $K$
is an effective killing rate reflecting the
balance of extinction and speciation at a given $z$, while
the proportionality  constant  $g$
determines what  fraction of the system is affected by
an evolutionary event.
Assuming that
evolution is the cause  of  extinction, we require that $K$ vanishes
if the evolutionary current  is zero.
Species going  extinct vacate a niche, which can be refilled at a later time.
In order to  formally account for this  flow in and out of
 the system,  it is convenient to introduce
 a `limbo' state, which absorbs extinct species,
and from which new species emerge at
 the  low fitness boundary  of the  system.
Requiring  a finite upper bound to   the total number of species
which the physical environment can sustain amounts to a
conservation law :  $ N(t) + \int_0^\infty P(y,t)dy = 1 $.
 In this notation
  $N(t)$ is   the fraction of species in the  limbo  state, while
$P(z,t)$ is the probability density of finding a living species with fitness
$z$.

The above  considerations result in
the differential   equations:
\begin{eqnarray}
\frac{ dN(t) }{dt} = -bN(t) + g \int_0^\infty P(y,t) K(P) dy
\label{PDE0}\\
\frac{ \partial P(z,t)}{\partial t} = -\frac{\partial (v(z)
P(z,t))}{\partial z} -
g P(z,t)  K(P)
\label{PDE}
\end{eqnarray}
where $b$ is the rate at which species are generated at the low fitness end
of the system.
The  initial and boundary  conditions are:
\begin{eqnarray}
N(t=0) = N_0 \\
\forall z: P(z,t=0) = P_0(z) \\
\forall t: \int_0^\infty P(y,t) dy = 1 - N(t) < \infty\\
\forall t: P(z=0,t) = bN(t)
\label{BC}
\end{eqnarray}
For later reference we  introduce the auxiliary function
\begin{equation}
D(t) = \log(t+1),
\label{deterministic}
\end{equation}
 which solves  the deterministic equation
of  a particle moving with velocity $v(z) = \exp(-z)$:
$d z/d t = \exp(-z)$, with   $D(0) = 0$.

A quantity often used to characterize  paleontological
data is the survivorship  curve of a cohort or the related
life span distribution\cite{Raup78}.
In our treatment the former quantity  corresponds to
the probability $W_t(\tau)$ that an agent
appearing at time $t$ will survive time $\tau$.
The latter can be  found
from $W_t(\tau)$ by differentiation:
\begin{equation}
R_b(\tau \mid t) = -\frac{d W_t(\tau)}{d\tau }.
\label{LIFESPAN}
\end{equation}
Here the subscript  stands for `born', to emphasize the
meaning of $t$.

In our model, an agent  surviving  time $\tau$
has  fitness $D(\tau) = \ln (\tau +1) $.
Since the  probability of being killed in the interval $d\tau$
is $K( P(D(\tau),t+\tau) ) d\tau$,
$W$ fulfills the differential equation:
\begin{equation}
\frac{d \ln W_t(\tau)}{d\tau } = - g K( P(D(\tau),t+\tau) );~\; \; \tau >0
\label{survival_prob}
\end{equation}
with initial condition $W_t(\tau = 0)=1$.
The limit of  $W$   for $\tau \rightarrow \infty$ is the probability
of eventually escaping extinction. Considering  that by far the largest
number of species which ever lived are now extinct \cite{Raup86}, we deem
model choices leading to a non zero ultimate survival  probability
to be irrelevant in an evolutionary context: Interesting cases require
$\int^\infty  K( P(D(\tau),t+\tau) ) d\tau  = \infty$.

Since, in general,  the time of appearance  of  species
is not known precisely, it is  of interest to  consider
the effect of averaging
 over a  time window $T$.
Weighting
$R_b(\tau \mid t)$  by  the normalized rate at which
new species flow into the system we obtain
the average
life-time distribution :
\begin{equation}
R(\tau) = \frac{\int_0^{T-\tau} N(t) R_b(\tau \mid t) dt}{\int_0^T N(t) dt}
\label{lifetime1}
\end{equation}
This averaging is non trivial,  if  the rate at which species appear
and  die changes in time.

Finally, the model extinction rate is simply the
 number  of species  per unit of time  which die at time $t$:
\begin{equation}
r(t)  = g \int_0^\infty P(z,t) K(P(x,t)) dx = dN/dt + bN
\label{ext_rate}
\end{equation}
Note that in the limit    $b \rightarrow \infty$,
$bN(t) \rightarrow r(t)$ describes the situation
considered  in Ref.~\cite{Sibani95},
 where    extinct species are immediately  replaced.

\subsection{Specific models }
To proceed with the mathematical analysis
we need to specify the killing term, which, as mentioned,   should 
1) only depend on the evolutionary current $v(z)P(z,t)$  
and 2) vanish when $vP$ vanishes. 
We  therefore   consider two variants of the model which are
consistent with these requirements:
 
\begin{description}
\item[a] The extinction rate depends on the total  rate of change
throughout the system:
 $ K = (\int_0^\infty -\partial(v P)/\partial y\; dy)^\alpha = P(0,t)^\alpha $.
 In this case the probability of an agent being killed is independent oflifetime
its fitness.

\item[b] The extinction rate depends on the rate of change {\em above} $z$:\\
$ K = (\int_z^\infty -\partial(v P)/\partial y \; dy)^\alpha  =
(v(z)P(z,t))^\alpha $.
This choice  similar in spirit to the model we have analyzed
 numerically.
The assumption makes
the interdependence of species asymmetrical: low-fitness species suffer
if  high fitness species evolve - but not vice versa. The older and fitter
an agent becomes, the  lesser is its probability per unit time of being killed.
\end{description}

In both  cases the parameter $\alpha$ is introduced because it
allows greater generality without unduly complicating the analysis.
It is meant  to  describe possible correlations  effects by
which  the size of the extinction cascade  following
the  evolutionary step  of one    species becomes  dependent  on  its
 fitness.
If $\alpha <1$ $ (>1)$, a move by an old,  slowly evolving  species
has a larger (smaller)  killing effect than
one by a  young, fast evolving species. In practice, good agreement
with data is obtained for $\alpha$ close to but  less than unity.

Model a) is at variance  with
empirical data. When contrasted  with the good results of model b)
this  lack of success  reveals the important role played by the
 asymmetry of the interactions.

\subsubsection{Model a}
The analysis is rather  brief  and  
mainly aimed at showing that   the model is not a viable
description.
Utilizing $P(z=0,t) = bN(t)$, we easily
find a differential equation for  $N$ alone:
\begin{equation}
\frac{dN}{dt}= -bN + g (bN)^\alpha ( 1-N ).
\label{m2}
\end{equation}

With $\tilde{N}(t) = (bN(t))^\alpha$ we obtain that
 the probability that a species born  at time $t$ will
survive time $\tau$ obeys:
\begin{equation}
\frac{d \ln W_t(\tau)}{d\tau} = -g \tilde{N}(\tau + t).
\label{survival}
\end{equation}

We note that Eq. \ref{m2} always  has  the stationary  solution $N=0$.
Furthermore, for $\alpha < 1$ it has an additional non-zero stationary
solution, $N_s$.
A simple stability analysis shows that for $\alpha >1$ the zero solution is
asymptotically  approached  as $\exp (- bt)$, while for
$\alpha <1$ this  solution
is unstable and   $N_s$  is stable.
It now  follows  that,  for $\alpha >1 $,
$\tilde{N}(t)$ is integrable with respect to time,  leading to a non-zero
ultimate survival  probability. As mentioned,  this behavior  does not
describe the   extinction statistics.
In the  case $\alpha < 1$ the right hand side of Eq. \ref{survival} is
asymptotically
a constant, the survival probability decays exponentially, and
the rates of extinction and of birth of new species approach  constants,
also in disagreement  with  the empirical data.

Let us finally consider the case $\alpha = 1$.
Let $N_0=N(t=0)$ be the initial value of $N$, and let
$$
C = \frac{N_0}{g N_0 + 1 -g} .
$$
The solution of Eq. \ref{m2} has, for $g \neq 1$,  the form
\begin{equation}
N(t) = \frac{C(1-g) e^{-(1-g) bt }}{1 - C g
e^{-(1-g) bt }},
\label{gamma_large}
\end{equation}
while for $g = 1$ the solution is
\begin{equation}
N(t) = \frac{N_0}{ bN_0 t + 1}.
\label{gamma_one}
\end{equation}
We also note that, if $g > 1$, then $N(t) \rightarrow \frac{g 
-1}{g}$
for $t\rightarrow \infty$, while if $g \leq 1$, $N(t) \rightarrow 0$ in the
same limit. In both cases the decay is exponential, and both cases can be
ruled out as irrelevant by the same arguments as used above.

In  the case $g = 1$,  $N(t) $ asymptotically approaches
$1/t$.
Solving  Eq. \ref{survival} for $W$ with $N$ given by Eq. \ref{gamma_one} we find
\begin{equation}
W_t(\tau) = \frac{bN_0 t +1}{bN_0 (t + \tau) + 1}.
\end{equation}
The average life-time distribution  is most easily obtained
by averaging  the conditional distribution
$R_b(\tau \mid t) = -dW_t(\tau)/d\tau$ over the rate at which
species appear, $bN(t)$.
 Omitting proportionality constants the result is
\begin{equation}
R(\tau) \propto \frac{1}{1+N_0b\tau}.
\end{equation}
Even though these last  choices  lead   to
 predictions which are not in strong disagreement
 with the data,  fine tuning two 
 parameters does not seem  acceptable  in the
lack of specific evidence.
This leads  us to conclude that
model a) is physically uninteresting.

\subsubsection{Model b} 
As a first step we multiply Eq. \ref{PDE} by $v(z)$, and obtain an 
equation for the new function $q = v P$. 
The solutions of this equation have the form $q(u(z,t))$, 
where $q$ is a function of just one 
variable
satisfying the non-linear - but separable - ODE
\begin{equation}
dq/du = -g q^{\alpha +1}
\label{q_eq}
\end{equation}
and where $u(z,t)$ satisfies the linear  inhomogeneous PDE  
\begin{equation}
\partial u/\partial t + v(z) \partial u / \partial z = 1.
\label{z_eq}
\end{equation}
The former equation  is solved by $q = (\alpha g u )^{-1/\alpha}$.  
To find the solution of the latter,  we let  $A$ and $B$ be  two  
arbitrary functions of a single real variable $z$, which
are   continuous for $z>0$ and which  vanish identically  for  $z<0$.
For $v = \exp(-z)$, the general solution has the form 
$u(z,t) = \epsilon \exp(z) + (1-\epsilon ) t + A(t+1-\exp(z))+ B(\exp(z) 
-(t+1))$
for some constant $\epsilon < 1$.
Thus,  the general  solution of Eq. \ref{PDE} has the  form 
\begin{equation}
P(z,t) = \frac{\exp(z)}
{\left[ g \alpha (\exp(z)-1) + A(t+1-\exp(z))+ B(\exp(z) 
-(t+1))\right]^{1/\alpha}}.
\label{sol1}
\end{equation}
Utilizing the  initial and boundary conditions, we  find
\begin{equation}
A(z) = (bN(z))^{-\alpha} , \; \; z>0
\end{equation}
and 
\begin{equation}
B(z) = (z+1)^\alpha P_0( \log(z+1))^{-\alpha} - g \alpha z,\; \; z>0. 
\end{equation}
 Equation  \ref{sol1} can now be reshuffled into
\begin{eqnarray}
P(z,t) = \frac{\exp(z)}
{\left[ g \alpha t +(\exp(z)-t)^\alpha P_0^{-\alpha}(\log(\exp(z) -t)) 
\right] ^{1/\alpha}};~ \; \; z > D(t)
\label{sol2a} \\
P(z,t) = \frac{\exp(z)}
{\left[ g \alpha (\exp(z)-1) + (bN(t+1 - \exp(z)))^{-\alpha} \right] 
^{1/\alpha}};~ \; \; z < D(t)
\label{sol2b}
\end{eqnarray}
Note that the solution is continuous in $z$, although its derivative will
in general be discontinuous at $z = D(t)$.

The survival probability of a species born at time $t$,
obtained by solving Eq. \ref{survival_prob},  is
\begin{equation}
W_t (\tau ) = \left[ 
\frac{(bN(t))^{-\alpha} }{(bN(t))^{-\alpha}+g \alpha \tau } 
\right]^{1/\alpha}.
\label{survive} 
\end{equation}
As  $W$  vanishes for large $\tau$  all  species   eventually go extinct.
 This happens here  for all  values of
$\alpha$, unlike the case previously described, where ultimate extinction with 
probability one is only achieved for  $\alpha < 1$. 
The above formula expresses a survivorship curve of a cohort\cite{Raup78},
and the corresponding distribution of life spans follows from it by
differentiation with respect of time (see Eq. \ref{LIFESPAN}).  If $\alpha $ is 
close to unity,
we find a $\tau^{-1}$ law for the former quantity, and
a $\tau^{-2}$ for the latter, which is in good agreement with paleontological
evidence. Further analysis of the model explains 
 what happens when life span distributions are averaged
over time dependent rates of extinctions and speciation.

For the sake of simplicity we 
only consider a limit case in which
all  probability is initially in the limbo state.
This situation can e.g. be achieved  by a limiting
procedure, where 1)  we  
choose  $P_0(z)= bC/( b+ C) \exp(-z C)$
and $N(t=0)=C/(C+b)$ for some $C$
(this fulfills the boundary conditions for any finite $C$),
and 2) we subsequently let  $C \rightarrow \infty$. 
In this limit  the expression for $P(z,t)$ in the region
 $z < D(t)$  remains  Eq. \ref{sol2b}, while $P$  vanishes identically
for $z >D(t)$.
 
A non-linear integral equation containing   $N(t)$ alone  is 
obtained by integration of    Eq. \ref{sol2b}, followed  by a change of 
variables.
The result is
\begin{equation}
1 - N(t) = \int_0^t \frac{ dy}{
\left[g \alpha (t-y) + (bN(y ))^{-\alpha} \right]^{1/\alpha} 
}. \label{eqforN}\\
\end{equation} 
Differentiating Eq. \ref{eqforN} with respect to time, and utilizing 
Eq. \ref{ext_rate}, we find 
the extinction rate: 
\begin{equation}
r(t) = \int_0^t \frac{g \; dy}{
\left[g \alpha (t-y) + (bN(y))^{-\alpha} \right]^{1 +1/\alpha } 
}. \label{Eqforr}\\
\end{equation}

The life time distribution  averaged over a 
window $T$ is found by using
Eq. \ref{sol2b}  in conjunction with Eqs. \ref{survival_prob}
and  \ref{lifetime1}. This  yields
\begin{equation}
R(\tau) = \frac{
  \int_\tau^{T }
 \frac{g dy}
      {\left[ g \alpha \tau + (bN(y))^{-\alpha} \right]^{1 + 1/\alpha 
}
      }
}
{\int_\tau^T  r(y) dy }.
\label{mellem1}
\end{equation} 
Equivalently, one can   average the 
life-time distribution of species born at time $t$ over the
normalized rate at which new species flow  into the system.
 This yields
\begin{equation}
R(\tau) = \frac{
  \int_0^{T-\tau }
 \frac{g dy}
      {\left[ g \alpha \tau + (bN(y))^{-\alpha} \right]^{1 + 1/\alpha 
}
      }
}
{\int_0^{T-\tau } bN(y) dy }.
\label{mellem2}
\end{equation}
Finally, in order to better describe the $\tau$ dependence
of $R$, it is useful to   express $N^{-\alpha}$ in terms of the survival
probability using Eq. \ref{survive}. The procedure yields:
\begin{equation}
R(\tau) = g (g \alpha \tau)^{-1-1/\alpha}
\frac{
  \int_\tau^{T }
 (1 - W_y^\alpha(\tau))^{1+1/\alpha} dy
}
{\int_\tau^{T} r(y) dy }.
\label{mellem3}
\end{equation}

The $\tau $ dependence stemming from  the limits of the
above integrals is only
relevant when $\tau \approx T$, and can be safely ignored,
if $T$ is sufficiently large. 
Furthermore, Eq.~ \ref{survive}
shows that, if $\alpha g \tau > (bN(y))^{-\alpha} $
is fulfilled throughout the integration interval,
then $W_y(\tau) \approx 0 $, and the 
$\tau$ dependence of the integrand becomes  negligible. 
In this case,  $R \propto \tau^{-1 - 1/\alpha}$,
i.e.\ an algebraic decay with an exponent
close to  $-2$ for  $\alpha \approx 1$.
As  shown numerically below, the assumption regarding
$r(t)$ is confirmed
by  model calculations for 
a wide range of parameter values.
In particular,
when  $\alpha $ close to unity the model yields
$r(t) \approx bN(t) \approx t^{-\delta}$ ,
with $\delta$ close to $0.5$ for a considerable time 
range. In this
situation, even though $\tau <<T$,
 $ \tau > T^{\delta \alpha}$  can be  fulfilled
for reasonable  $\tau$ values. Consider for instance
 $\delta = 0.5$, $\alpha = 1$ and 
 $T = 10^{12}$  (generations). The power law behavior
with exponent $-2$ then occurs for 
 $\tau > 3 \times 10^4$ (generations) which is rather short
life-time.  
The  cross-over from a  slow  decay to   
the  $\tau^{-2}$ law  is  also found in simulations
of the full model, where  moves to shorter times
 as the degree of interaction among species increases,
and the decay exponent $\delta$ correspondingly  decreases.
(see e.g. Figs.~5 and 6).

 From the point of view of biological data
assuming that  the extinction rate varies slowly
 over some typical time scale  of species life time, 
seems very reasonable: The  rate of extinction appears
to change  appreciably
 on a scale of a  hundred  million  years, while
the  range of species life-times is better expressed in
millions of  years.

Returning to the analysis of the model,
we first  note a major difference
in the asymptotic behavior for
$\alpha <1$ and $\alpha \geq 1 $.
In both cases the time independent  function $P_\infty(z)$ 
obtained  by taking 
$t \rightarrow \infty$  and by setting $N(t) = N_\infty \neq 0$ in 
Eq. \ref{sol2b} formally satisfies the model equations.  
However, only for  $\alpha < 1$ is  $P_\infty(z)$ normalizable and
thus  a true solution. The corresponding steady state value of  $N$, 
  $N_\infty$  is then implicitly given by the relation
$1-N_\infty = (bN_\infty)^{1-\alpha}/(g(1-\alpha))$, which
always has a solution in the unit interval. 

In the case 
$\alpha \geq 1$,  normalizability of $P(z,t)$ 
requires that   $N(t) \rightarrow 0 $ for $t \rightarrow \infty$.
No steady state solution can then exists, since 
$P(z,t)$ vanishes with $t$ at any fixed $z$,
as e.g. in the familiar case of simple diffusion on the infinite line.

Even when a steady state solution exists
the fact that  $D(t)$ only changes logarithmically
shows, in connection with
Eqs. \ref{sol2a} and  \ref{sol2b}, that the relaxation
is  extremely  slow, and that,
depending on the initial condition, the transient
behavior might well be the only relevant one.

The long time asymptotic solution  can be obtained explicitly 
in the case $\alpha = 1$.  If $bt >> 1$   the term
$dN/dt$ in Eq. \ref{PDE0} is negligible, and 
 $bN(t) \approx r(t)$.
In this limit  we can  also  neglect $N$ compared to one, thus  
finding the following  approximate equation for the rate of extinction:
\begin{equation}
1 = \int_0^t \frac{g dy}{g (t-y) + r(y)^{-1}},
\end{equation}
which  is solved by $r(t) = (g t)^{-1}$.

\begin{figure}[t]
\centerline{\psfig{figure=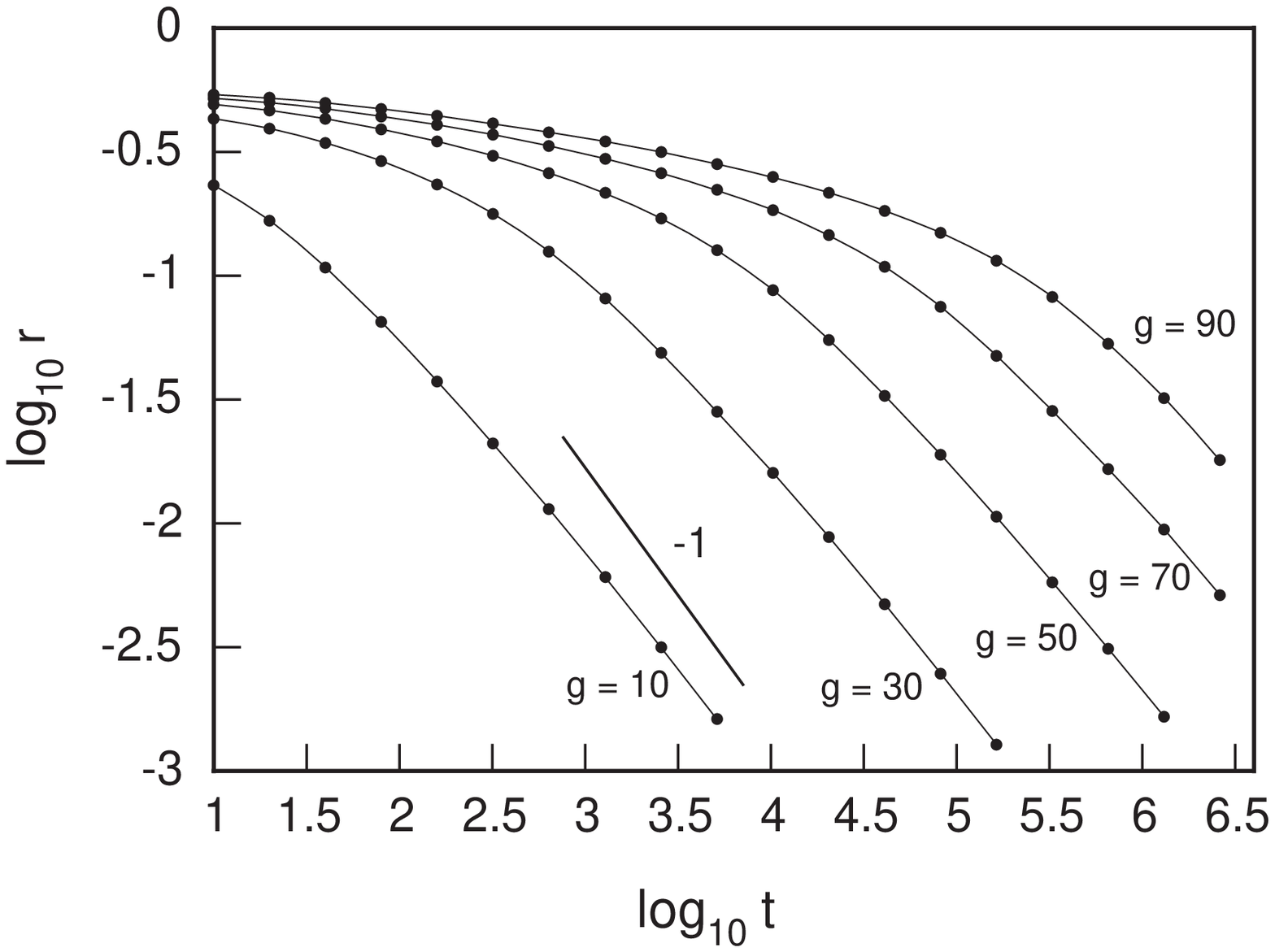,height=10cm,width=14cm}}
\begin{caption} {}
\begin{scriptsize} 
We show the logarithm of rate of extinction of the analytical model plotted
as a function of the logarithm of time, for $\alpha = 1$, $b= 1$ and various values
of the coupling constant $g$. Note the change from a slow to a faster
decay, and the fact that the cross-over time between the two regimes increases with $g$.
This behavior strongly resembles that of the full reset model shown
in Fig.~4. 
\end{scriptsize}
\end{caption} 
\end{figure}
                  
In Fig.~9 we show the  the rate of extinction of the
mean field model as a function of time,  for $\alpha =1$,
$b=1$  and several
choices of the coupling constant $g$. Note the striking 
similarity with the  extinction rate of the 
full  reset model,  shown in Fig.~4. In both cases 
there is a cross-over from a power-law with a
(numerically) small effective decay exponent
 to to another power-law with exponent approaching $-1$. Furthermore, 
the cross-over time strongly increases with the degree
of coupling, which is expressed in one case by
thethe parameter $CR$  and in the other by $g$.
  
\section{Summary and conclusion}
In this paper we have described a model of evolutionary dynamics
which builds on two basic elements: 1) the   dynamics of a 
single species in a constant physical environment is assumed
to  follow  a record statistics,
leading to a logarithmic improvement of different evolutionary measures, and
2) extinctions are caused by co-evolution: the evolution of a species removes 
its less fit neighbors. This approach  is analyzed  numerically as well as
analytically in a simpler, mean field version.

Our first model assumption stems  from considering
the behavior of random walks in highly dimensional
rugged fitness landscapes. It 
 stresses the fact that
a species maintains a  record of its past history
through a stable genetic pool and states that the
statistical properties of this pool can only  change 
when a random   mutation appears,  which is better
than the  `best so far'. Its predictions are in good     
agreement with the  intersting empirical data 
obtained by Lenski and Travisano \cite{Lenski94} on the evolution 
of   bacterial cultures in a nutrient deprived  environment. 
 Our second assumption   leads to good   agreement
 with    empirical data regarding 
 the  life time distribution of species and  the decay of   
 the extinction rate. The size distribution of events, which is the main focus
 of  many  other (stationary) approaches, is in our case derived from the
 the    rate of extinction, as discussed already in the introduction. Also
 in this respect  our model is in good  agreement with the data.
 The behavior of the system as a whole is 
  characterized by the fact that, as time goes, the
  dynamics approaches that of a single agent optimizing its
  fitness. In this sense one could talk about increasing correlations
  among the species. When the system is partly randomized by 
  external   catastrophes modeling sudden changes of the
  environment, the whole system is set back in its evolution as shown by 
  the strong rebound effect in the rate of extinction.
 Interestingly, catastrophes do not   change
the life-time distribution in any  significant way.
 In summary,  our model offers a 
single, albeit approximate, explanation of most evolutionary data, linking 
the distribution of life spans with the behavior of the rate of
extinction and of the distribution of event sizes. 
   
From the vantage point of a 
theoretical physicist, the scale invariance present
in evolution and extinction data
 begs for an explanation. 
As mentioned in  our introductory discussion
 several  different  theoretical approaches
 currently exist. 
  On the opposite  extreme of the 
modeling spectrum,  purely {\em statistical} 
considerations have lead to the suggestion
that   extinctions are  basically due to external 
periodic forcing, stemming  impacts from  from celestial bodies\cite{Raup84,Sepkoski89}.
As biological evolution is a non-reproducible experiment (within
the time scales available to human observers ), the dilemma
of what really has happened might not be easily solved.  
Biological history   contains a good measure of frozen
 accidents which do not  have and probably do 
 not  require  specific explanations. Therefore,
 it  seems hard (and pointless)  to  
 exclude that  different  mechanisms
for evolution and extinction (i.e. rocks from the sky {\em and} species
competition) could have  play together to shape the course  of evolution.     
We would expect  progress in clarifying  the relative  importance
of various  modeling elements  to   come from 
 closer scrutiny of the  basic assumptions, particularly
 at the level of population dynamics.  
 
{\bf Acknowledgments}\\             
P. S. would like to thank the  Santa Fe Institute of Complex Studies
for nice hospitality and Mark Newman for interesting discussions.
This work was supported in part by Statens   Naturvidenskabelige Forskningsr\aa d.


\begin{thebibliography}{10}
\bibitem{Raup_disagrees} We note that in more recent works
 Raup  criticizes the   separation in `background' and `mass' extinctions
 as artificial. See e.g.
Ref.~\cite{Raup95}. Our data analysis (i.e. the values of the 
exponents)  depends very weakly on making this assumption. 
However, at least within the theoretical model we propose, the distinction
between `mass' and `background' is   meaningful and corresponds to two different
modes of extinction, one endogenous and the other exogenous.  Our model predicts the
decaying background extinctions  and robustly absorbes externally
imposed mass extinctions.
\end{thebibliography}
\end{document}